\documentclass[twocolumn]{aastex631}

\usepackage{xcolor}
\usepackage{amsmath}
\usepackage{float}

\shorttitle{Underground muons}
\shortauthors{Fedynitch, Woodley, Piro}

\graphicspath{{./}{figures/}}

\newcommand{\sibyll}[1]{{\sc sibyll#1}}
\newcommand{\ddm}{{\sc ddm}}
\newcommand{\dpmjet}[1]{{\sc dpmjet#1}}
\newcommand{\qgsjet}{{\sc qgsjet}}
\newcommand{\eposlhc}{{\sc epos-lhc}}

\newcommand{\mceq}{{\sc mceq}}
\newcommand{\bartol}{{\it Bartol}}
\newcommand{\proposal}{{\sc proposal}}
\newcommand{\mute}{{\sc mute}}
\newcommand{\rmd}{{\mathrm d}}

\newcommand{\MUSIC}{{\sc music}}
\newcommand{\MUSUN}{{\sc musun}}

\begin{document}

\title{On the Accuracy of Underground Muon Intensity Calculations}

\author{A. Fedynitch}
\email{anatoli@gate.sinica.edu.tw}
\affiliation{Institute for Cosmic Ray Research, the University of Tokyo,
5-1-5 Kashiwa-no-ha, Kashiwa, Chiba 277-8582, Japan}
\affiliation{Institute of Physics, Academia Sinica, Taipei, 11529, Taiwan}
\author{W. Woodley}
\email{wwoodley@ualberta.ca}
\author{M.-C. Piro}
\email{mariecci@ualberta.ca}
\affiliation{Department of Physics, University of Alberta, Edmonton, Alberta T6G 2E1, Canada}
\date{\today}

\begin{abstract}

Cosmic ray muons detected by deep underground and underwater detectors have served as an information source on the high-energy cosmic ray spectrum and hadronic interactions in air showers for almost a century. The theoretical interest in underground muons has nearly faded away because space-borne experiments probe the cosmic ray spectrum more directly, and accelerators provide precise measurements of hadron yields. However, underground muons probe unique hadron interaction energies and phase space, which are still inaccessible to present accelerator experiments. The cosmic ray nucleon energies reach the hundred-TeV and PeV ranges, which are barely accessible with space-borne experiments. Our new calculation combines two modern computational tools: \mceq{} for surface muon fluxes, and \proposal{} for underground transport. We demonstrate excellent agreement with measurements of cosmic ray muon intensities underground within estimated errors. Beyond that, the precision of historical data turns out to be significantly smaller than our error estimates. This result shows that the sources of high-energy atmospheric lepton flux uncertainties at the surface or underground can be significantly constrained without taking more data or building new detectors. The reduction of uncertainties can be expected to impact data analyses at large-volume neutrino telescopes and for the design of future ton-scale direct dark matter detectors.

\end{abstract}

\section{Introduction} \label{sec:intro}

For more than 80 years, it has been known that muons observed at the surface or by experiments underground reflect the properties of high-energy cosmic rays interacting with the Earth's atmosphere \citep{Barrett:1952woo, Bugaev:1998bi,Kochanov:2008pt, Gaisser:2016uoy,pdg2020}. As the topic evolved over many decades, the theoretical activity focussed on searches of prompt atmospheric muons from decays of charm or heavier flavour mesons and the characterisation of the cosmic ray composition at energies around the knee of the cosmic ray spectrum \citep{Elbert:1982xj, Bugaev:1989we, Hoerandel:2002yg, BeckerTjus:2020xzg}. 

One of the latest theoretical analyses of underground muon intensities was performed 20 years ago by Bugaev, Misaki, and Naumov (BMN) \citep{Bugaev:1998bi} to characterise the prompt flux and to search for its signature in data. With the advent of large-volume neutrino telescopes, the searches have been extended from facilities underground to those underwater and under ice \citep{Gaisser:1994yf, Gondolo:1995fq, Sinegovskaya:2000bv}. All attempts to detect the flux of prompt muons or neutrinos remain unsuccessful, and only a few extreme models have been excluded, \textit{e.g.}\ by LVD \citep{LVD:1999khf} or recently by IceCube \citep{IceCube:2015wro, IceCube:2016umi}. These current results are in good agreement with the conclusions of BMN.

Another field with an interest in accurate deep underground muon calculations is direct detection dark matter (DM) experiments. Cosmic ray muons are a significant background for DM searches (see, \textit{e.g.}\ \cite{Formaggio:2004mm} for a review), producing secondary neutrons in interactions with rock or surrounding materials. These muon-induced neutrons pose challenges for the direct detection of DM nuclear recoils since they can mimic the expected energy spectrum from WIMP-nucleus (Weakly Interacting Massive Particles) scattering. By installing DM experiments in deep underground laboratories, the overburden acts as a natural shield, resulting in a suppression of muon-induced backgrounds by several orders of magnitude. To some extent, current large-scale DM detectors rely on underground muon calculations to estimate muon-induced neutron fluxes and find the most effective shielding strategies. A well-characterised muon flux estimation is essential for the planning of future detectors where data-driven background estimates are not yet available. A phenomenological approach by \citet{Mei:2005gm} (MH) used an empirical fitting formula to find a consistent description of data for the characterisation of experimental backgrounds without drawing a connection to the cascade calculations. 

For high-energy atmospheric lepton flux calculations, the historical underground muon data are highly relevant, considering that the uncertainty for fluxes or ratios of TeV-range muons and neutrinos is surprisingly large. Neutrino fluxes are known within 40\%, while errors for particle ratios exceed this value \citep{Barr:2006it, Honda:2006qj, Fedynitch:2012fs}. This is attributed to the experimental challenges involved in measuring TeV muons at the surface with sufficient accuracy to obtain data-driven constraints, and to the lack of high-energy accelerator data taken within a relevant phase space \citep{Honda:2019ymh,Fedynitch:2021ks}. Direct measurements of TeV muon spectra have been performed at near-horizontal zenith angles by DEIS \citep{Allkofer:1985ey} and MUTRON \citep{Matsuno:1984kq}, topping out at a few TeV with large uncertainties. TeV muons have lost most of their energy during transport and can be observed at lower energies deep underground, giving indirect constraints on the surface fluxes. However, the uncertainty that stems from the ``unfolding'' of the surface fluxes comes with additional penalties that render the high accuracy of underground data less useful than it could be.

The reduction of atmospheric neutrino uncertainties is crucial for data analyses that systematically depend on the correctness of the neutrino flux prediction, such as measurements of the neutrino cross-sections \citep{IceCube:2017roe}, searches for sterile neutrinos \citep{IceCube:2020phf}, earth tomography \citep{2012AGUFM.P23D..03H,Donini:2018tsg}, and astrophysical neutrino measurements using neutrino-induced muon tracks \citep{IceCube:2015wro,Stettner:2019tok}. Underground muons are an accessible, data-driven source of unique constraints on the uncertainties that cannot be obtained otherwise.

On the other hand, the accuracy of muon transport simulations and of computational methods has significantly progressed. Several Monte Carlo codes are available for muon transport and intensity simulations, \textit{e.g.} {\sc propmu}, based on calculations by \citet{Lipari:1991ut}; {\sc mum} \citep{Sokalski:2000nb}, based on the BMN calculation; and \MUSIC{} and \MUSUN{} \citep{Kudryavtsev:2008qh}, based on a parameterisation of LVD data \citep{LVD:1998lir}. General-purpose particle transport codes such as {\sc fluka} \citep{Battistoni:2015epi} or {\sc geant4} \citep{GEANT4:2002zbu} can also be used for muon transport. Dedicated muon transport codes that do not contain a model for surface fluxes include \textit{e.g.} {\sc mmc} \citep{Chirkin:2004hz}. The most recent of these codes is \proposal{} \citep{koehne2013proposal}, which we employ in our study. For calculations of surface fluxes, most approaches use analytical formulae, such as those described in \citet{Gaisser:2016uoy}, or numerical results, as in BMN; we use the newer code \mceq{} \citep{Fedynitch:2018cbl}, with more details provided in Section~\ref{sec:surface_models}.

We are revisiting this topic because the original questions remain, namely: what is the contribution of the prompt flux to high-energy neutrinos, and how degenerate is it with the astrophysical flux \citep{IceCube:2016umi,IceCube:2020wum}? Is there an intrinsic charm component to the prompt flux \citep{Brodsky:1980pb,Gondolo:1995fq}? What is the reason for the quantitative discrepancy between calculations and measurements of the inclusive muon flux at the surface \citep{Lagutin:2004ka,Fedynitch:2012fs,Fedynitch:2018vfe,Fedynitch:2021ks}? Is it possible to reduce the uncertainties of the high-energy atmospheric neutrino calculations below 40\% \citep{Barr:2006it,Fedynitch:2012fs,Sinegovskaya:2014pia,Evans:2016obt}? Can the spectrum and composition of light cosmic ray nuclei be constrained above 100 TeV/nucleon, yielding tighter constraints on the composition around the knee \citep{Mascaretti:2019mnk}? And can the systematic shifts observed between space-borne direct cosmic ray detectors be resolved \citep{AMS:2015tnn,CALET:2019bmh,DAMPE:2019gys}?

In this work, we aim to cross-check the accuracy of the modern computational tools \mceq{} and \proposal{} at TeV and multi-TeV energies. We estimate the potential to constrain high-energy atmospheric neutrino calculations by applying the uncertainty estimation schemes used for atmospheric neutrinos to muons, and we propagate the uncertainty to the underground intensities. We believe that the analysis of hadronic uncertainties for underground intensities has not been previously performed. Our computational scheme is made publicly available as the tool \mute{} (\textbf{MU}on In\textbf{T}ensity Cod\textbf{E}), such that our results can be easily reproduced and used for further studies.

\section{Muons from cosmic ray interactions} \label{sec:surface_models}

\subsection{Surface} \label{MCEq}
\begin{figure}
\centering
\noindent\includegraphics[width=.95\columnwidth]{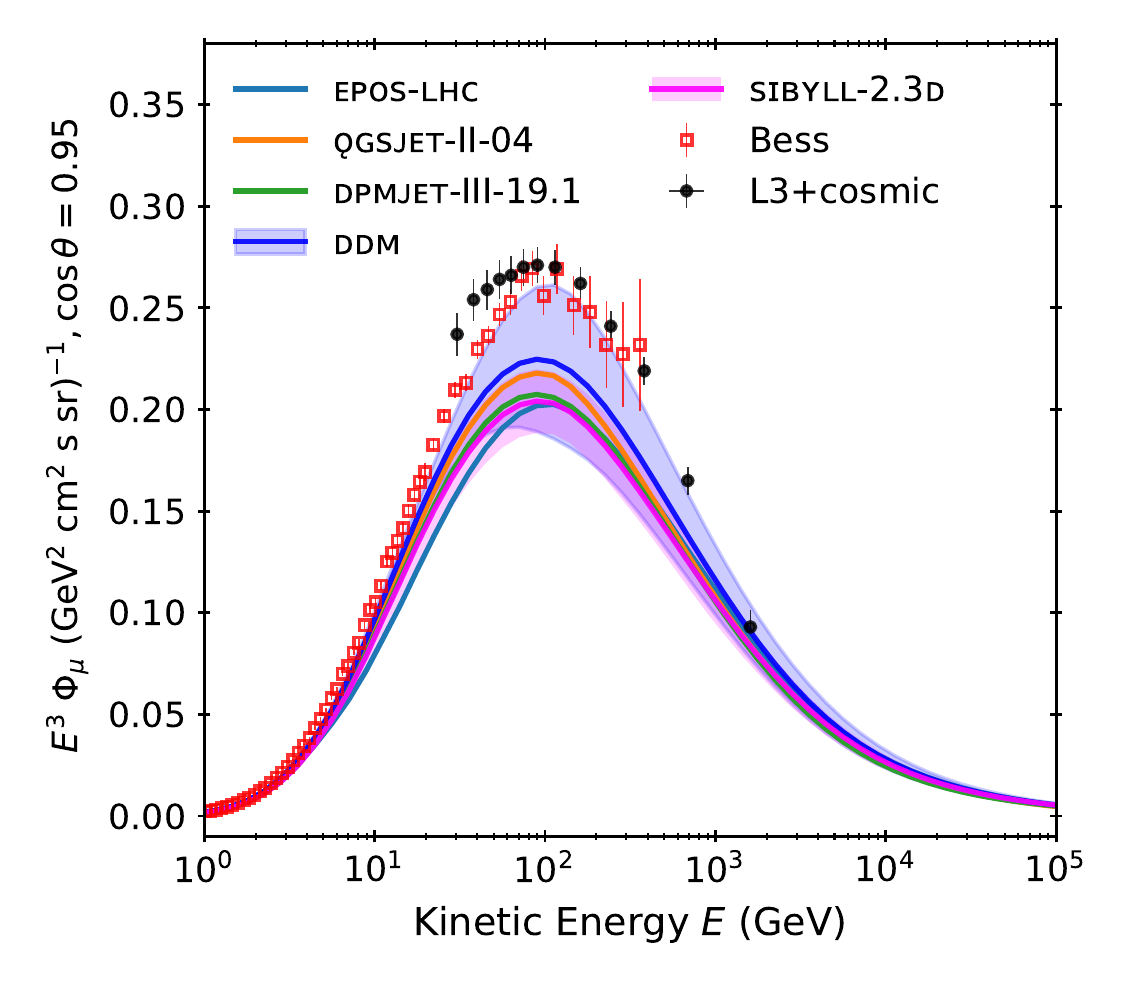}
\noindent\includegraphics[width=.95\columnwidth]{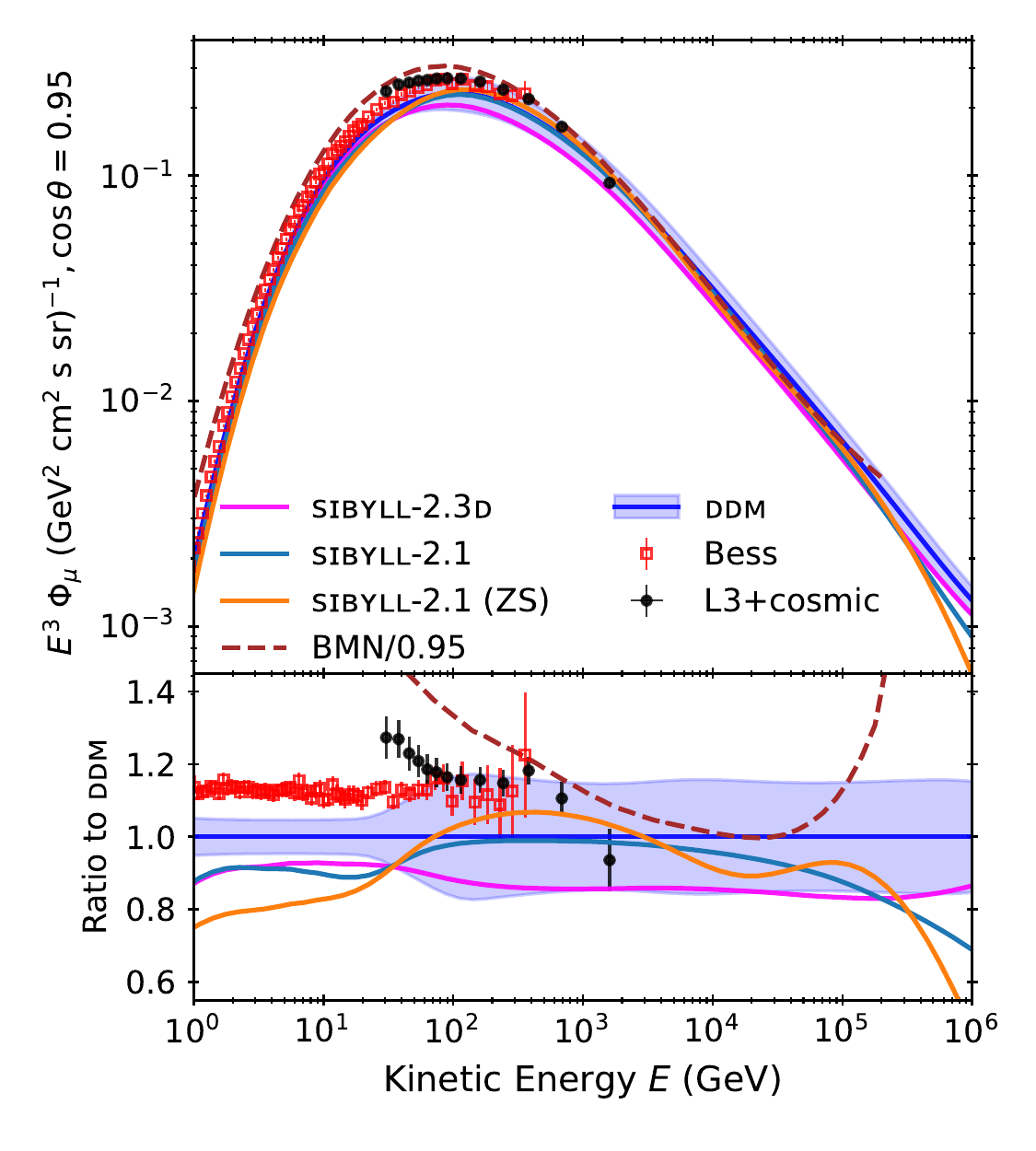}
\caption{Vertical muon intensity from \mceq{} compared to spectrometer data at near-vertical zenith angle of $\cos(\theta) = 0.95$, on linear (upper panel) and logarithmic (lower panel) scales. The cosmic ray flux model is the Global Spline Fit (GSF) \citep{Dembinski:2017zsh} and the atmosphere is the US \citet{us_std_atmosphere}. The uncertainty bands represent hadronic uncertainties, which dominate at energies below 1 TeV. At higher energies, the cosmic ray flux uncertainty is an additional source of $\sim25\%$ uncertainty. Systematic uncertainties are not shown for BESS and L3+cosmic, which can partly compensate for the offset between \ddm{} and the data.\label{fig:vertical_surf_flux}}
\end{figure}

The hadronic interactions of cosmic rays with the atmosphere create secondary hadrons, such as charged pions and kaons, which decay into more stable particles when moving towards the ground. Atmospheric (cosmic) muons and neutrinos are the most abundant species at sea level. Their fluxes have been characterised by calculations based on different techniques \citep{Volkova:1983yf,Kochanov:2008pt,Gaisser:2016uoy,Fedynitch:2012fs, Fedynitch:2018cbl,Kochanov:2019yvx}. Since they interact very feebly, they are also the most deeply penetrating. Other secondaries are absorbed within the first few metres of rock.

We use the program \mceq{} \citep{Fedynitch:2015kcn, Fedynitch:2018cbl} to solve the one-dimensional cascade equations and calculate the fluxes of muons at the surface for different zenith angles and combinations of particle production, atmospheric density, and primary cosmic ray flux models. \mceq{} accounts for realistic density profiles of the atmosphere, particle interactions, decays, and energy losses. It achieves this with an accuracy comparable to or exceeding established cosmic ray air shower simulators, such as {\sc corsika} \citep{Heck:1998vt,Engel:2018akg}, {\sc conex} \citep{Bergmann:2006yz}, and {\sc aires} \citep{Sciutto:1999jh}, and within a small fraction of their computational time. Recent comparisons between Monte Carlo simulations and \mceq{} can be found in, \textit{e.g.}~\citet{Gaisser:2019xlw,CORSIKA8:20212c,Kozynets:2021LK}.

The vertical muon flux prediction from \mceq{} is shown in Figure \ref{fig:vertical_surf_flux}. The Monte Carlo event generators \dpmjet{-III-19.1} \citep{dpmjetIII,Fedynitch:2015kcn}, \qgsjet{-II-04} \citep{Ostapchenko:2010vb}, \eposlhc{} \citep{Pierog:2013ria}, and \sibyll{-2.3d} \citep{Engel:2019dsg, Fedynitch:2018vfe} are currently the most accurate particle interaction models for cosmic ray cascades in air. 

The Data Driven interaction Model (\ddm{}) \citep{Fedynitch:2021ks} is a new inclusive model designed for high-precision atmospheric lepton flux calculations. In contrast to the event generators, \ddm{} is accompanied by uncertainties (blue band) derived from fitting the model to data from fixed-target experiments at accelerators. \citet{Barr:2006it} estimate uncertainties from the availability and phase space coverage of measurements dated before the release of relevant proton-carbon data by NA49 \citep{Alt:2006fr,Anticic:2010yg} and NA61 \citep{Abgrall:2015hmv}. An implementation of this \bartol{} error scheme based on \mceq{} is shown as the magenta band around \sibyll{-2.3d} in Figure \ref{fig:vertical_surf_flux}. Alternatively, uncertainties have been estimated by fitting surface muon data in \citet{Honda:2006qj} and \citet{Sanuki:2006yd}, and more recently by \citet{Honda:2019ymh}, and \citet{Yanez:2019bnw} and \citet{Yanez:2021qd}.

The fluxes from current event generators are confined by the magenta band. The upper edge of the error band is 10-15\% lower than the data without accounting for systematic uncertainties. Despite being directly constrained by accelerator data, the central expectation from \ddm{} also falls short of describing surface muon fluxes but it is consistent within uncertainties. The lower panel of Figure \ref{fig:vertical_surf_flux} shows remarkable agreement between the combination of \ddm{} and the GSF model between the spectral index and the Bess data as a constant 10--12\% offset. Although the origin of this offset is, at the moment, unknown and is not relevant for underground fluxes, we speculate that it is related to systematic uncertainties of the primary spectrum that were not accounted for in GSF or the data. The combination of the deprecated model \sibyll{-2.1} \citep{Ahn:2009wx} and the Zatsepin-Sokolskaya (ZS) primary spectrum \citep{Zatsepin:2006ci} has been found by \citet{Kochanov:2008pt} to be in good agreement with the surface muon spectrum. This combination also falls short to match the surface muon spectrum in particular at lower energies. However, at TeV energies, the models favoured by \citet{Kochanov:2008pt}, \sibyll{-2.1} and the BMN calculation, are well-contained by the uncertainty band of \ddm{}. In the following, we concentrate on the newest models \sibyll{-2.3d} + \bartol{} uncertainties and the \ddm{}, since they well represent the differences in muon fluxes between the variety of models.

The choice of the cosmic ray flux model for the spectrum of nucleons at the top of the atmosphere also has a significant impact on the surface fluxes. Global Spline Fit \citep{Dembinski:2017zsh} is fixed as the cosmic ray flux model choice in this work. As discussed in Appendix \ref{app:primary_flux_and_hardonic_models}, none of the following conclusions are significantly affected by this choice.

\subsection{Underground}

\begin{figure}
\centering
\noindent\includegraphics[width=\columnwidth]{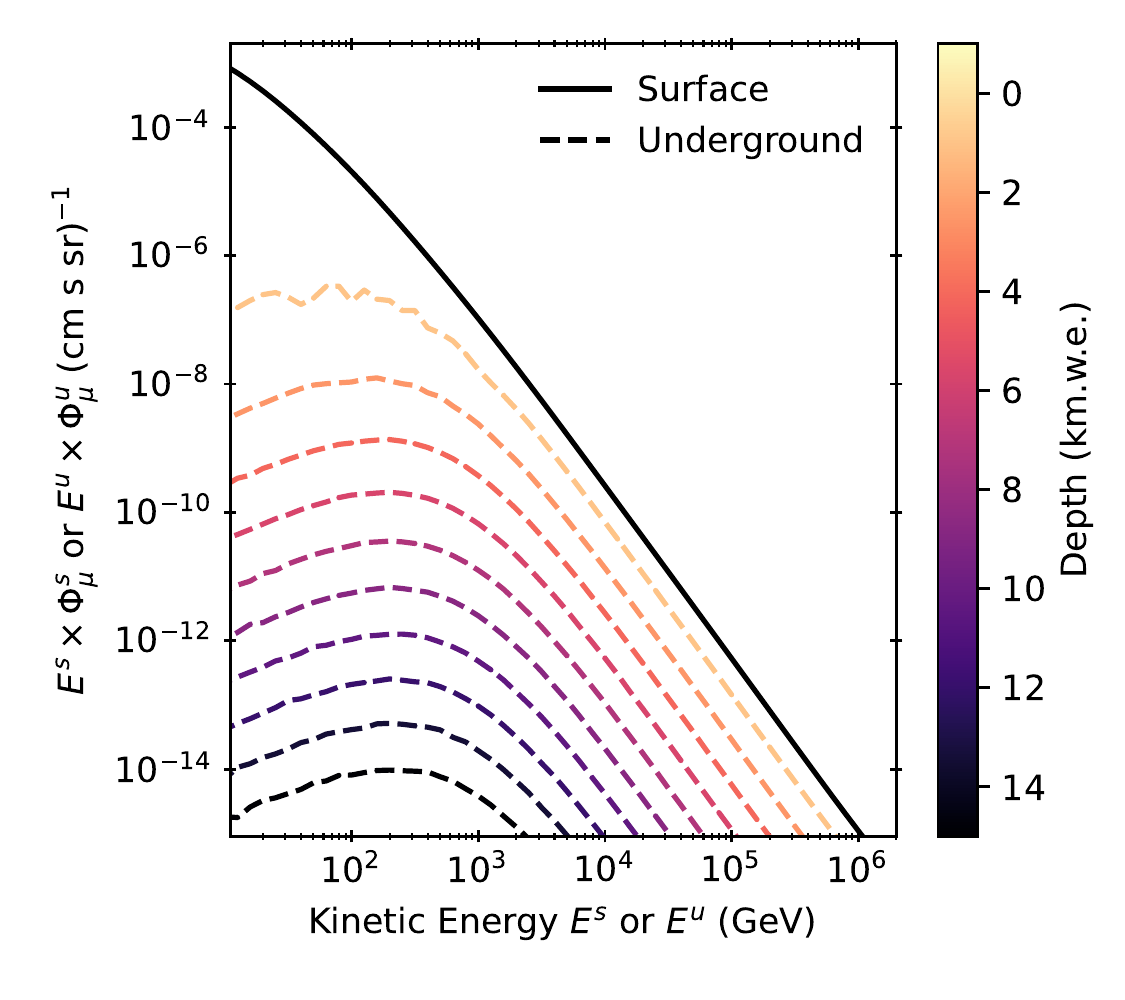}
\caption{Vertical muon spectra at different depths in rock weighted with energy, calculated with \mceq{} using \sibyll{-2.3d} and GSF cosmic ray flux. The propagation underground has been simulated with the \proposal{} propagation code \citep{koehne2013proposal,dunsch_2018_proposal_improvements}. The stopping of low-energy muons suppresses fluxes at lower energies, shifting the mean energy (the peak) successively to higher energies. \label{fig:underground_fluxes}}
\end{figure}

\begin{figure}
\centering
\noindent\includegraphics[width=\columnwidth]{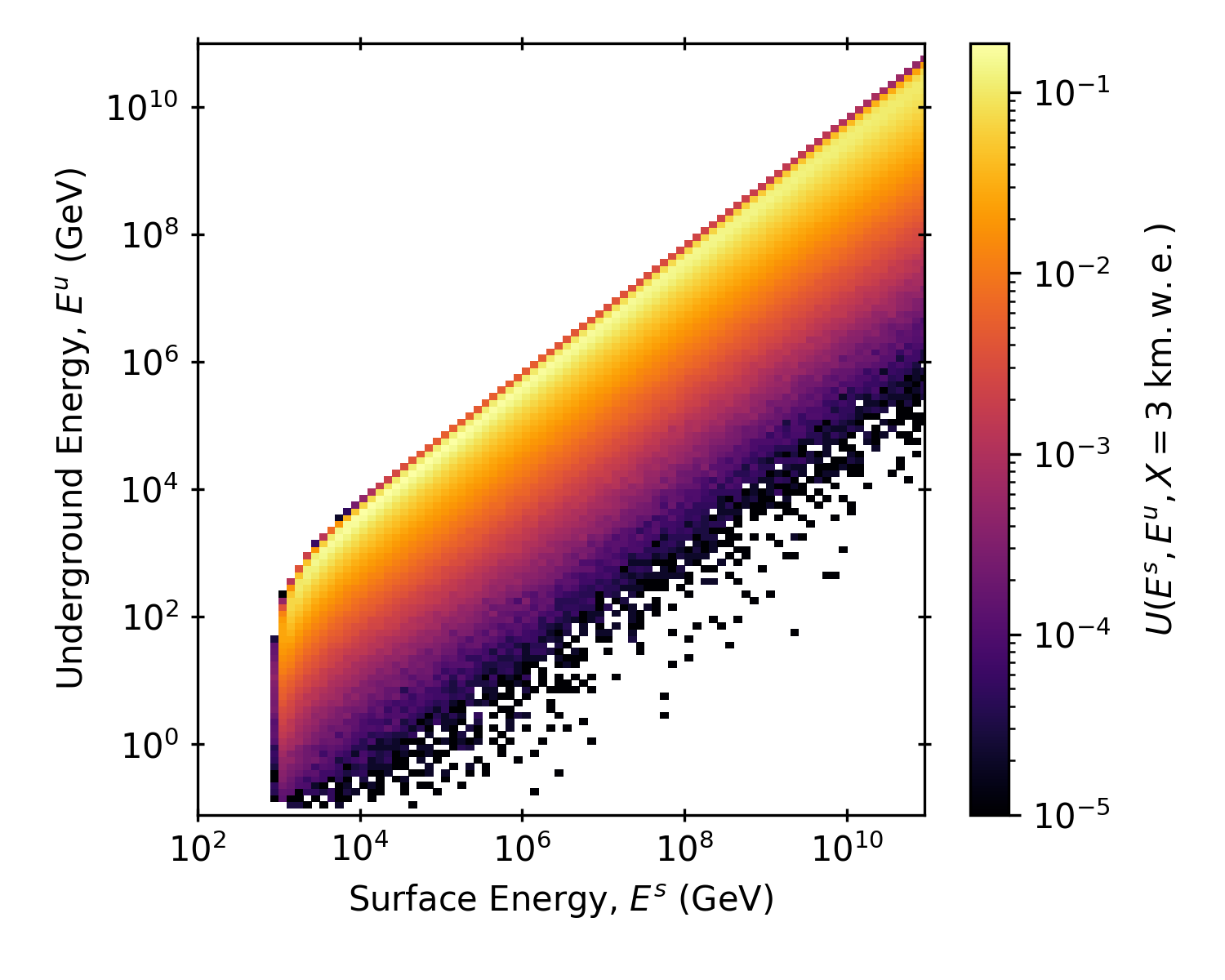}
\caption{Surface-to-underground transfer matrix for a depth of 3 km.w.e., simulated with \proposal{} for $10^5$ initial muons per surface energy-angle bin. The colours scale according to the survival probabilities as defined in Section \ref{PROPOSAL}. \label{fig:transfer_matrix}}
\end{figure}

\begin{figure}
\centering
\noindent\includegraphics[width=.95\columnwidth]{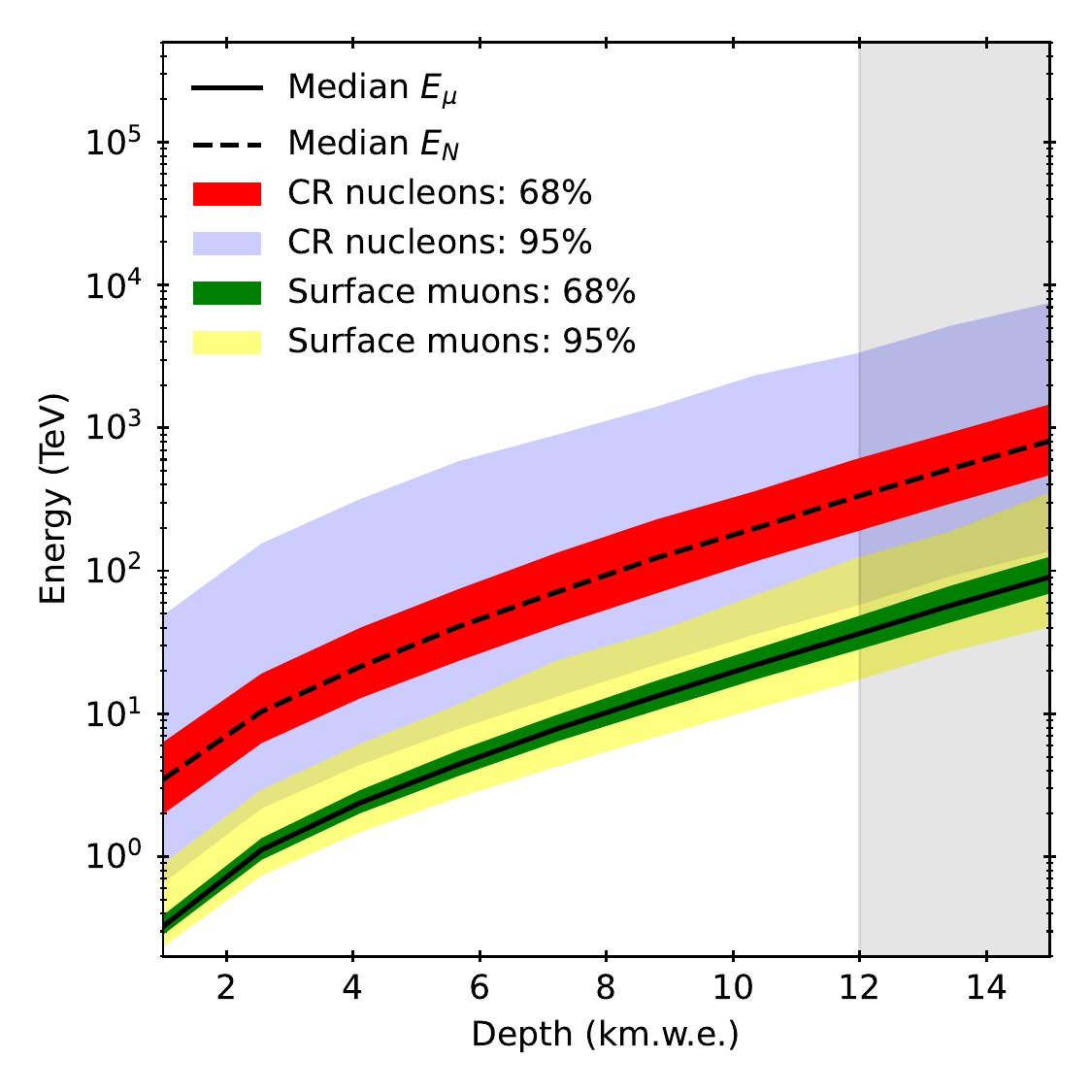}
\caption{Energies of surface and cosmic ray (CR) nucleons contributing to the muon rate underground, shown for the \sibyll{-2.3d} hadronic interaction model and the GSF primary flux model. The coloured bands show the energy ranges that result in 68\% and 95\% of the underground muon rate respectively. At depths larger than 12 km.w.e. the rate becomes dominated by neutrino-induced muons through charged-current interactions as indicated by the grey band. \label{fig:equivalent_surface_cr_energies}}
\end{figure}

Once muons enter the ground, the leading energy loss processes are ionisation and pair production. In dense media, such as rock and water, most muons stop, capture electrons, form muonium, or undergo deep-inelastic scattering with rock nuclei accompanied by neutron emission (see \textit{e.g.}~\citet{Mei:2005gm, Manukovsky:2016fcn}). In air, most GeV muons decay. As illustrated in Figure \ref{fig:underground_fluxes}, lower-energy muons are successively absorbed while the mean underground muon energy increases. The loss of energy between surface and underground at different depths can be represented by a transfer tensor. An example for a depth of 1 km.w.e., as produced by \mute{}, is given in Figure \ref{fig:transfer_matrix}. The stochastic nature of radiative (``catastrophic'') losses at $E_{\mu}\gg 100$ GeV results in a smearing of the relation between surface and underground energy, as shown in this figure.

The relevant ranges of muon surface energies and initial cosmic ray energies are constrained by calculating the contribution of each surface or cosmic ray energy bin to the underground intensity (see Figure \ref{fig:equivalent_surface_cr_energies}). The fluctuations smear out the muon surface energy band, covering almost one decade. Due to the contribution from neutrino-induced muons above 12 km.w.e.,\footnote{Although cosmic muons at larger slant depths can potentially be separated from neutrino-induced muons, the arguments would become more detector-specific, which exceeds the general scope of this work.} the maximal surface energies that can be probed with cosmic underground muons range from several tens of TeV with a tail extending up to 200 TeV. Cosmic ray nucleon energies cover a range wider than a decade in energy due to the cascading process (see \cite{Barr:2006it, Sanuki:2006yd, Fedynitch:2018vfe, Fedynitch:2021ks} for more detail).

\begin{figure}
\centering
\noindent\includegraphics[width=.95\columnwidth]{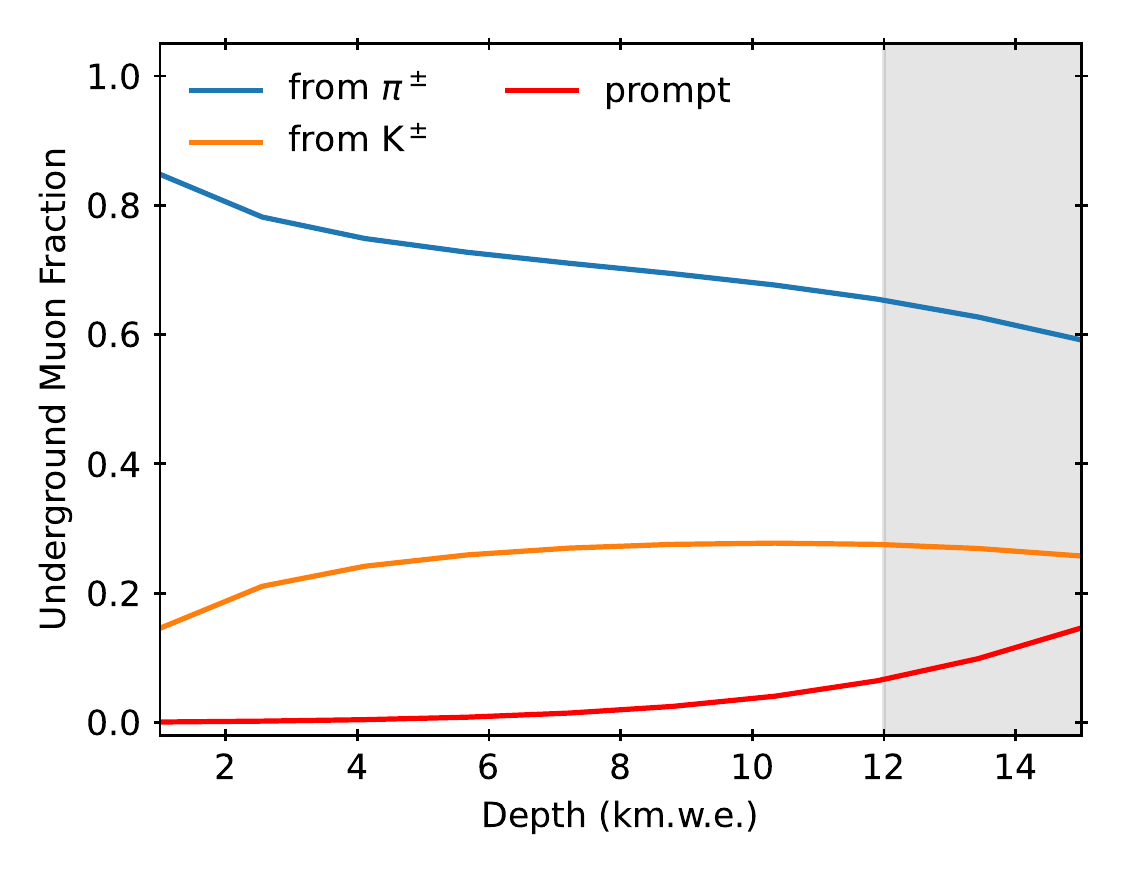}
\caption{Fraction of parent mesons contributing to the vertical-equivalent underground intensity of muons, calculated with \sibyll{-2.3d} and GSF primary flux. Since deep underground muons sample TeV surface energies, the kaon contribution is at its maximum of $\sim20\%$. The charm model in \sibyll{-2.3d} yields a $\sim10\%$ prompt muon component below depths where neutrino-induced muons contribute (grey band). \label{fig:meson_fractions}}
\end{figure}

The surface energies are therefore too low to probe the prompt muon component from decays of short-lived charm or unflavoured mesons \citep{Bugaev:1998bi,Garzelli:2015psa,Illana:2010gh,Fedynitch:2018cbl}, which dominates over the conventional muons from pion and kaon decays at surface energies above PeV. As shown in Figure \ref{fig:meson_fractions}, the prompt flux from the \sibyll{-2.3d} model stays within 10\% over the entire accessible depth range. In combination with a post-LHC hadronic interaction model, we confirm that the conclusions of the analysis by BMN, which found no evidence for a prompt component in underground muon data, are unchanged.

Primary cosmic ray energies exceed PeV only in the tails and at the largest of depths. Due to the small statistics of experiments at very large km.w.e., probing the mass composition at the knee of cosmic ray spectrum with deep underground muon intensity measurements is infeasible except for the case of a very early cutoff in the proton spectrum. On the other hand, quantitative constraints should be possible for the energy range at the transition from direct to indirect cosmic ray measurements between 100 TeV and PeV.

\section{Computational scheme}

Our calculation is a convolution of surface flux tables from \mceq{} with surface-to-underground transfer tensors computed with \proposal{}, a successor program of the {\sc mmc} code developed originally for the AMANDA experiment at the South Pole by \citet{Chirkin:2004hz}. In the following, the underground flux and intensity computations are described in more detail.

\subsection{Propagation of Muons Underground} \label{PROPOSAL}

To simulate the propagation of muons through rock, we are interested in computing transfer tensors $U(E^s, E^u, X)$ that associate a muon energy at the surface, $E^s$, with a distribution of underground energies, $E^u$, at a slant depth
\begin{equation}
\label{eq:slant_depth}
    X=\rho_\text{rock}\frac{h}{\cos(\theta)},
\end{equation}
where $\rho_\text{rock}$ is the rock density in g cm$^{-3}$, $h$ is the vertical depth in cm, and $\theta$ is zenith angle at the surface. The depth in water equivalent units $\rho_\text{water} X$ is a convenience unit that simplifies the comparison of underground rates, which become independent of the overburden density.

For the \proposal{} simulations, the geometry used is a large sphere of homogeneous, isotropic standard rock ($Z=11$, $A=22$, $\rho=2.65$ g cm$^{-3}$). Muons are initially placed at $(0, 0, 0)$, the center of the sphere, and are propagated downwards for initial energies corresponding to the center values of the \mceq{} energy grid. The surface fluxes obey azimuthal symmetry in the relevant energy range of $E^s$. This reduces the number of angular bins in the \proposal{} calculations to a set of zenith angles within $[0,\ 90)^{\circ}$, where 0 describes downward-going muons, and $90^{\circ}$ describes horizontal muons. Although the distinction between surface zenith angle and underground zenith angle is possible to be made in \proposal{}, because the average angular deviation from surface to underground is $\left<\left|\theta^u-\theta^s\right|\right><1^{\circ}$, the distinction is not drawn, and so the simulations are only one-dimensional, and for computational efficiency, scattering is turned off. Additionally, both $\mu^-$ and $\mu^+$ can be defined and propagated in \proposal{}; however, because the energy loss in the rock is almost the same for both polarities, only $\mu^-$ is propagated for simplicity.

\proposal{} propagates the muons one-by-one in a Monte Carlo simulation until the muon decays, loses all of its energy, or has reached the end of the set propagation length. Muons can lose energy via multiple different processes, the cross-sections for which are described by different theoretical models in \proposal{}. For these simulations, the Bethe-Bloch-Rossi model \citep{Rossi:99081} is used to describe ionisation, the Kelner-Kokoulin-Petrukhin models are used for electron-positron pair production \citep{Kelner:1998mh} and bremsstrahlung \citep{Kelner:1995hu}, and the updated 1997 parameterisation by Abramowicz, Levin, Levy, and Maor \citep{ABRAMOWICZ1991465,Abramowicz:1997ms} is used for photonuclear interactions. While the accuracy of the models is not perfect and some cross-sections remain uncertain, we expect a negligibly small impact on the underground intensity calculation, at most a few percent \citep{Kokoulin:1999bn,Sokalski:2002dk,Sandrock:2020zaf}. However, the uncertainties remain relevant for the simulation of stochastic photon emission patterns during the transport of very high-energy muons in dense media.

If a muon reaches the depth $X$ without decaying, its underground energy, $E^u$, is filled into a histogram $\rmd N/\rmd E^u$, which is then normalised by the number of simulated events (typically $10^5$). These histograms are arranged as columns of $U(E^s, E^u, X)$. The simulations are repeated for each surface energy bin $E^s$ that enumerates the columns of $U(E^s, E^u, X)$. A two-dimensional image of a slice of the transfer tensor $U(E^s, E^u, X=3~\mathrm{km.w.e.})$ is shown in Figure \ref{fig:transfer_matrix}.

The survival probability $P(E^s, X)$ for a muon with surface energy $E^s$ to reach a certain depth $X$ can easily be obtained by integrating over the underground energies:
\begin{equation}
\label{eq:survival_probability}
P(E^s, X)=\int_0^\infty U(E^s, E^u, X)\mathrm{d}E^u.
\end{equation}
\begin{figure}

\centering
\noindent\includegraphics[width=\columnwidth]{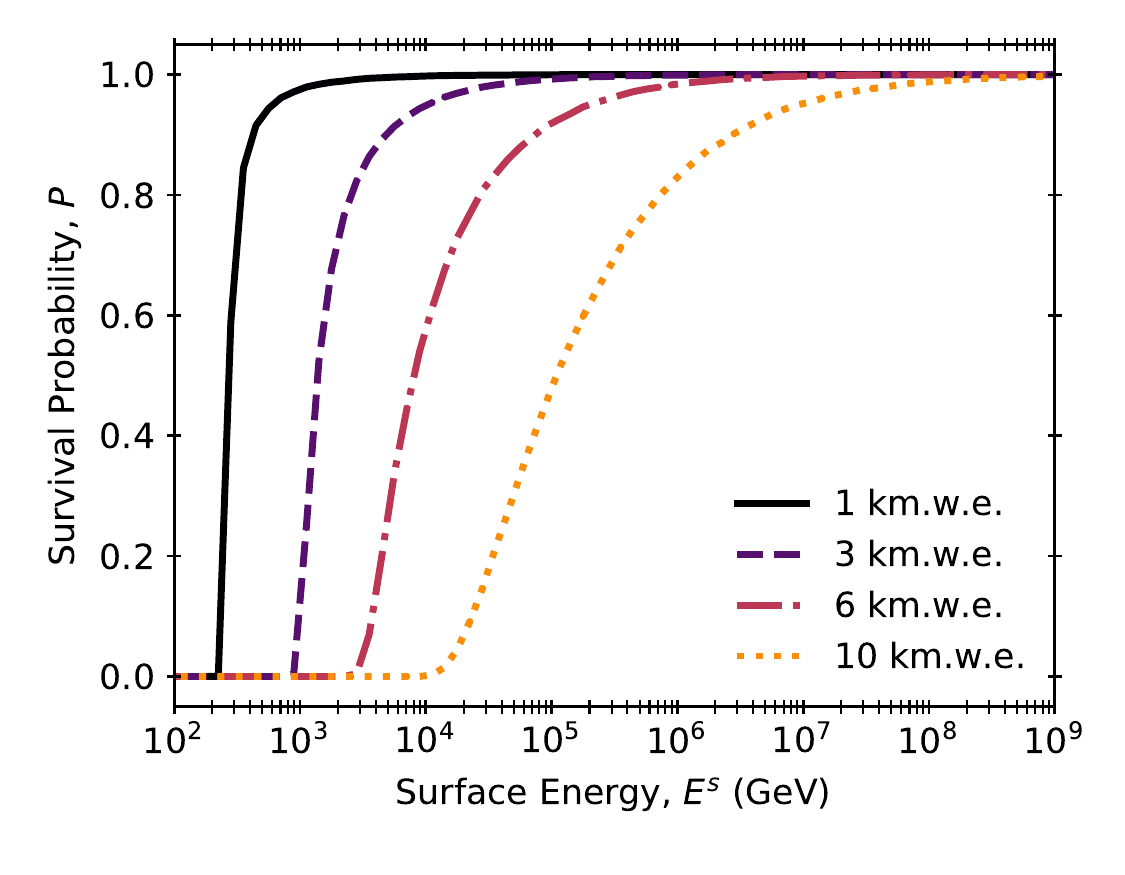}
\caption{Survival probability against surface energy for various slant depths. \label{fig:survival_probability}}
\end{figure}

\noindent The integrals over the columns of the transfer matrices, such as the one illustrated in Figure \ref{fig:transfer_matrix}, are shown as survival probabilities in Figure \ref{fig:survival_probability} for multiple values of $X$. As surface energy increases, the survival probability increases, as muons with higher energies have more of a chance to survive the propagation. Considering individual high-energy muons, their survival depends more on the number of discrete catastrophic loss events and less on the average stopping power $\langle \rmd E/\rmd X\rangle$, as it is the case for ionisation dominated losses. In this high-energy regime, the tails, or the smearing in energy, which are simulated by \proposal{} or other particle transport codes, are important and can not be treated as continuous losses without sacrificing accuracy.
 
\subsection{Underground Fluxes}

The underground muon flux, $\Phi^u(E^u, X, \theta)$, is calculated through a convolution of the surface flux $\Phi^s(E^s, \theta)$ obtained from \mceq{} and the transfer tensor $U(E^s, E^u, X)$:
\begin{equation}
\begin{split}
\label{eq:convolution}
    \Phi^u(&E^u_j, X_k, \theta_k)=\\
        &=\sum_i\Phi^s(E^s_i, \theta_k)U(E^s_i, E^u_j, X_k)\Delta E^s_i
\end{split}
\end{equation}
\noindent where $X$ is the depth along the initial trajectory and depends on the zenith angle, as defined in Eq.~(\ref{eq:slant_depth}). This discrete version of the convolution illustrates better than integrals that the underground fluxes are the result of a simple algebraic operation.

\subsection{Vertical Underground Intensity and Zenith Angle Dependence}
\label{sec:intensity}

Integrating the flux, defined by Eq.~(\ref{eq:convolution}) over the underground energy gives the underground intensity, $I^u(X, \theta)$, as a function of slant depth and zenith angle:
\begin{equation}
\label{eq:intensity}
I^u(X, \theta)=\int^{\infty}_{E_{th}}\Phi^u(E^u, X, \theta)\mathrm{d}E^u,
\end{equation}
where $E_{th}$ is a detector's energy threshold, here set to $0$ GeV for generality. For deeper $X$, the dependence on $E_{th}$ is small, since the mean muon energy increases with depth.

\begin{figure}
\centering
\noindent\includegraphics[width=\columnwidth]{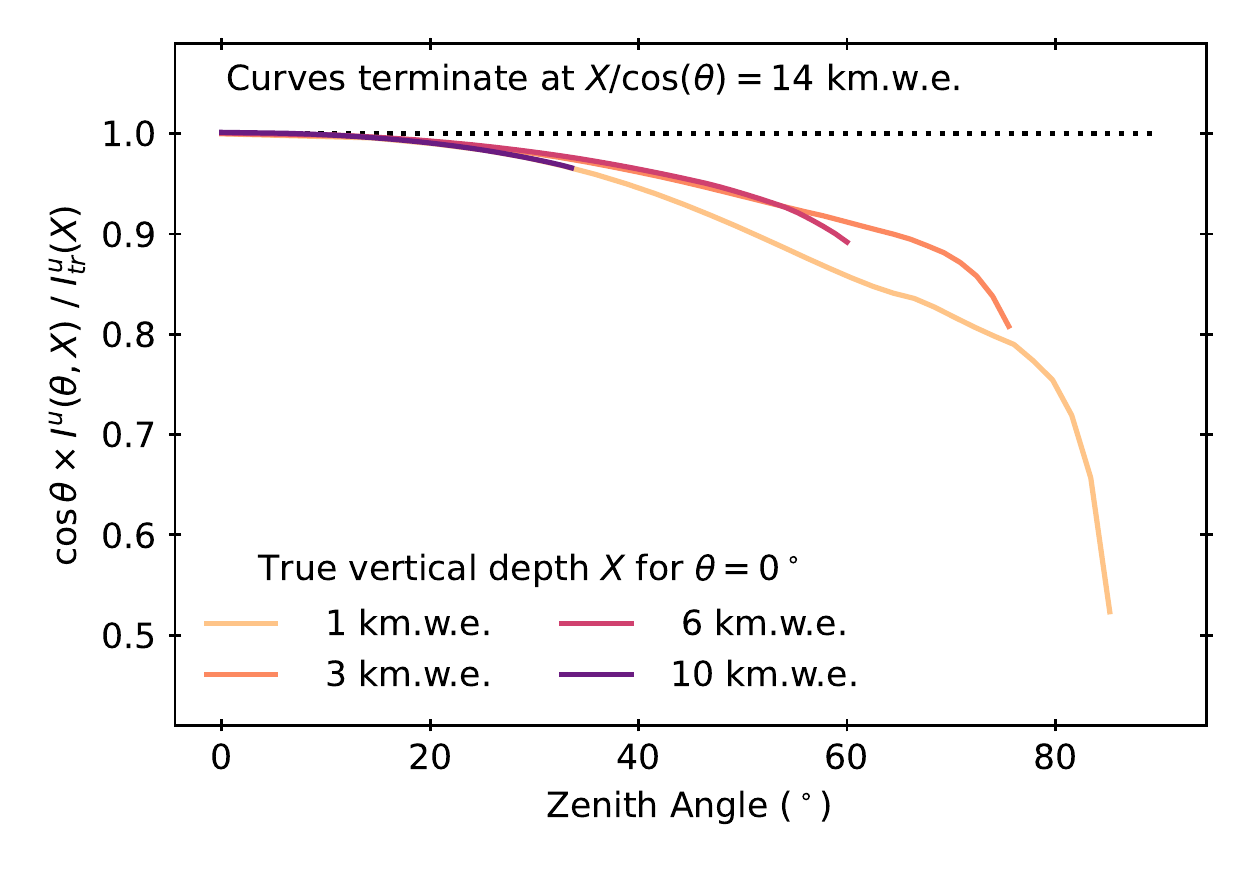}
\caption{The ratio of vertical-equivalent to true vertical muon flux as a function of the cosine of the zenith angle. The black dotted line is the expectation for the case of perfect agreement. The solid curves show deviation due to the approximation in Eq.~(\ref{eq:intensity_eq}) from the true vertical intensity. The approximation performs worse at larger zenith angles and shallower depths. \label{fig:deviation}}
\end{figure}

Multiplying the muon intensity by $\cos(\theta)$ gives the vertical-equivalent muon intensity,
\begin{equation}
\label{eq:intensity_eq}
    I^u_\text{eq}(X) = I^u(X, \theta)\cos(\theta).
\end{equation}
\noindent This is the form in which muon intensities are typically given in the literature since detectors are located at a fixed vertical depth. As it is widely known, this relation is only approximate. The deviation between the true vertical intensity, defined as
\begin{equation}
\label{eq:true_vert_intensity}
    I^u_\text{tr}(X)=I^u(X, \theta=0),
\end{equation}
and the vertical-equivalent intensity becomes significant as $\theta$ increases, which can be seen in Figure \ref{fig:deviation}. To avoid any bias or degeneracy with the cosmic ray spectra index, Eq.~(\ref{eq:intensity_eq}) should not be used for zenith angles significantly larger than $20^{\circ}$. While this aspect has been discussed in classic literature, some experimental results are still reported under the assumption of the $1/\cos(\theta)$ approximation at much larger zenith angles. Further discussion on how this affects the interpretation of experimental data is given in Appendix \ref{app:vertical_equivalent_approximation}. Since the true vertical intensity is typically inaccessible for laboratories located at a fixed geometrical depth, reporting double-differential measurements in $X$ and true zenith angle, $\theta$, would be beneficial for the theoretical analysis of the data as, for example, in \citet{LVD:1999khf}.

\subsection{\mute{}}

Our computational scheme is wrapped in a new code called \mute{} \citep{william_woodley_2021_5803936}.\footnote{The source code for \mute{} is publicly available at \url{https://github.com/wjwoodley/mute}.} It is a flexible, lightweight, open-source {\sc python} code, and, as described, uses \mceq{} (v1.2.2) and \proposal{} (v7.1.1) to calculate surface fluxes and survival probabilities respectively. The convolution described in Eq.~\ref{eq:convolution} allows the surface and underground geometries to be changed independently without repeating the entire calculation. The surface fluxes for many of the standard models as well as high-statistic ($N=10^5$) transfer tensors for standard rock and water are provided as data files supplied with the code, resulting in computation times for most underground and underwater flux and intensity calculations being on the order of seconds. With the most recent theoretical models available for use, \mute{} can provide forward predictions with the current lowest possible theoretical uncertainties.

\section{Results and discussion}

\begin{figure}
\centering
\noindent\includegraphics[width=\columnwidth]{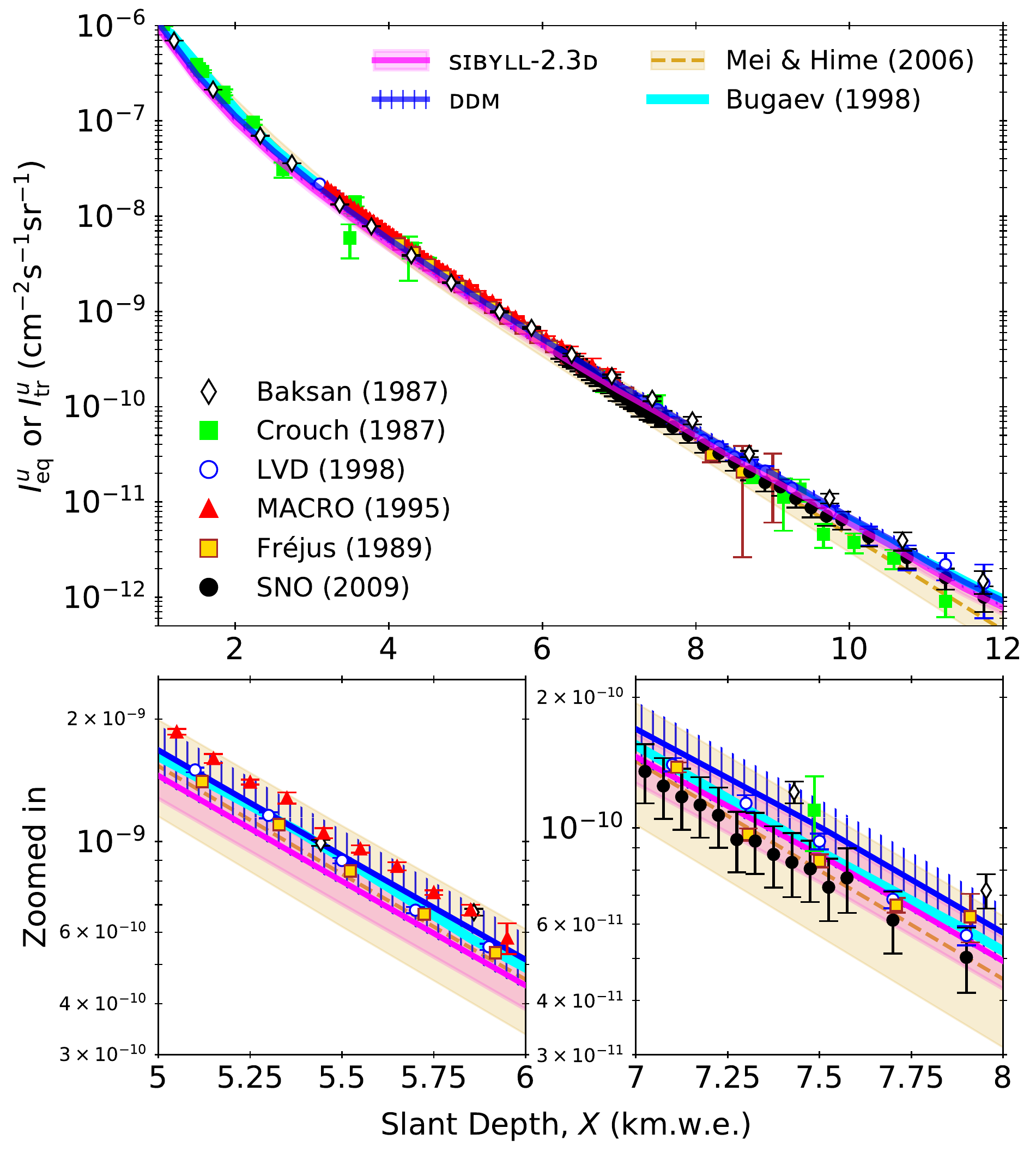}
\caption{True vertical underground muon intensity calculated for standard rock with the \sibyll{-2.3d} and \ddm{} models against slant depth, with experimental data from Baksan \citep{baksan1987}, Crouch \citep{crouch1987}, LVD \citep{LVD:1998lir}, MACRO \citep{macro1995}, Fréjus \citep{frejus1989}, and SNO \citep{sno2009}. Data and MH are vertical-equivalent, whereas the BMN and our calculations are for true vertical intensity. The ratio plot to \ddm{} in Figure~\ref{fig:comparison} reveals further detail. \label{fig:intensity_true}}
\end{figure}

\begin{figure}
\centering
\noindent\includegraphics[width=\columnwidth]{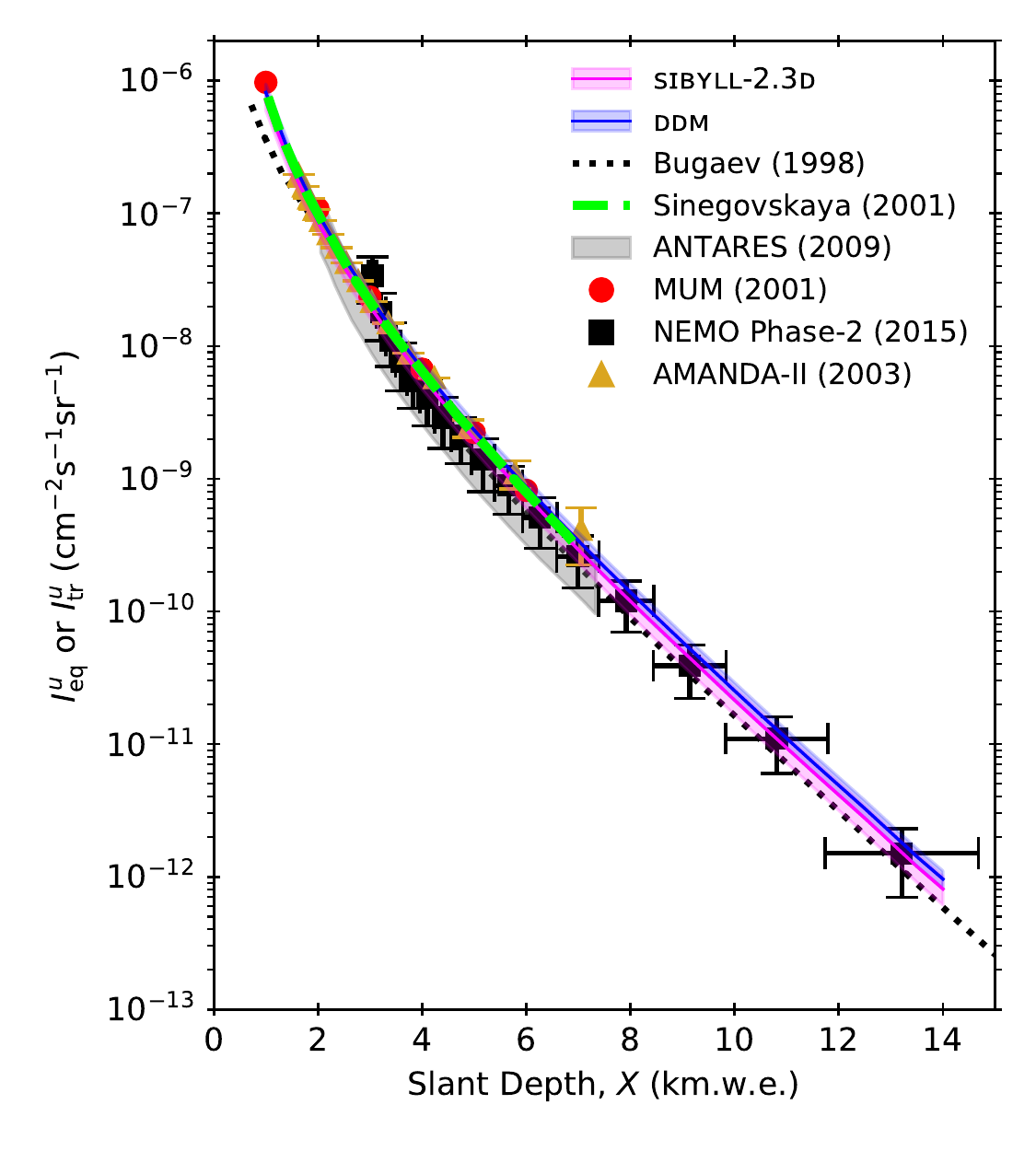}
\caption{True vertical underground muon intensity calculated for water with the \sibyll{-2.3d} and \ddm{} models against slant depth, with experimental data from ANTARES \citep{Bazzotti:2009sy}, NEMO Phase-2 \citep{NEMO:2014htj}, and AMANDA-II \citep{Bai:2003mn}, calculations from BMN and Sinegovskaya \citep{Sinegovskaya:2000bv}, and simulation from MUM \citep{Sokalski:2000nb}.} \label{fig:intensity_water}
\end{figure}

\begin{figure*}
\centering
\noindent\includegraphics[width=\textwidth]{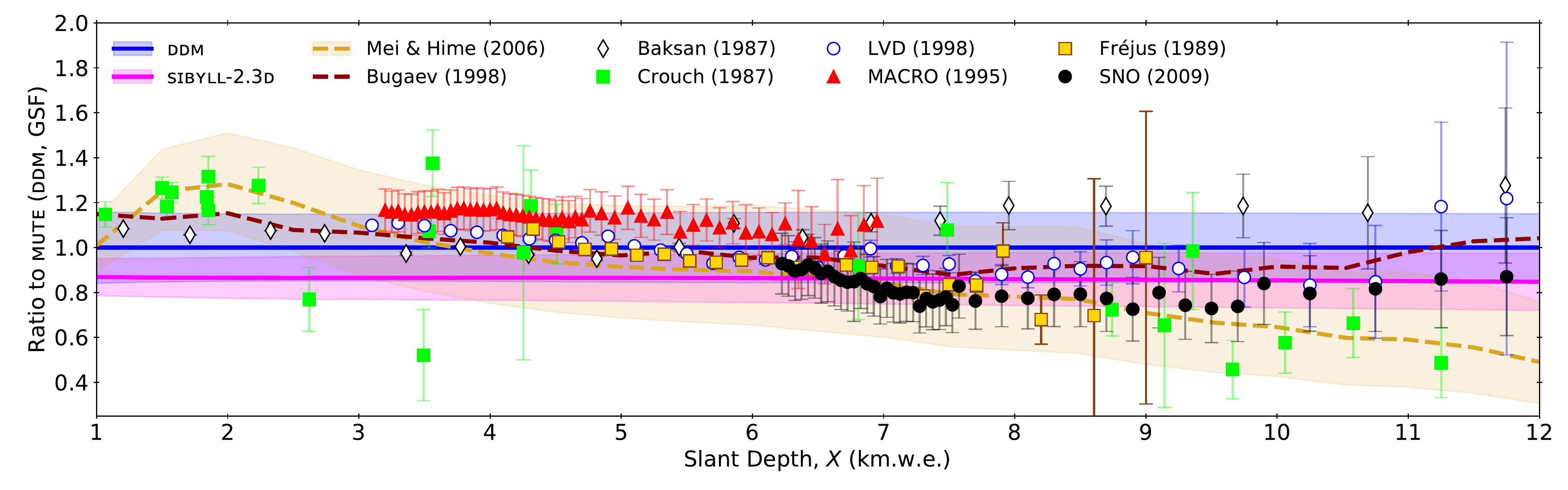}
\caption{The ratio to \mute{} results for standard rock using \ddm{} and GSF of the experimental data and the predictions made with \sibyll{-2.3d} with uncertainties computed using the \bartol{} method, the MH fit, and the BMN calculation. All experimental data are referenced in Figure \ref{fig:intensity_true}. Systematic and statistical errors are geometrically summed for LVD, MACRO, and SNO. Other error bars denote solely statistical errors. \label{fig:comparison}}
\end{figure*}

The true vertical underground muon intensity from Eq.~(\ref{eq:true_vert_intensity}) as calculated with \mute{} is plotted against slant depth in Figure \ref{fig:intensity_true} for standard rock, and in Figure \ref{fig:intensity_water} for water. There is excellent agreement between our results and the experimental data over the entire depth range within uncertainties, which can be seen in higher detail in Figure \ref{fig:comparison}. \ddm{} more accurately describes data taken at shallower slant depths, roughly corresponding to surface energies below 1 TeV, whereas \sibyll{-2.3d} is better for deeper slant depths that sample surface energies from TeV to tens of TeV.

Visually, the data align well with the fit by MH. However, the depth dependence of individual data sets follows that of our calculation more consistently. MH has a clear tendency to reproduce the classic \citet{crouch1987} ``world survey data set'' since it is based on the same fitting function. These data are a compilation of measurements from different laboratories and are accompanied by systematic uncertainties that are difficult to control. The difference in the slope between the yellow MH band and our calculation is a result of the simple parameterisation by a sum of two exponential functions. The relatively large errors of MH are estimated from the fit to underground data and can exhibit bias from systematic uncertainties and the fixed form of the parameterisation.

The BMN calculation uses inputs derived from data collected and analysed in the 1970s for conventional muon fluxes. The hadronic model is based on measurements at the CERN ISR and a predecessor framework of \qgsjet{}. The cosmic ray ``NSU'' model is an empirical parameterisation of data of that era \citep{Nikolsky:1984emk}. The data, without corrections for systematics, are in good agreement with this calculation. As later shown by \citet{Lagutin:2006kv} and \citet{Kochanov:2008pt}, this agreement can be attributed to the surface muon spectrum at TeV energies, as also shown in Figure \ref{fig:vertical_surf_flux}. Given the quality and age of the data it was derived from, the flux must be accompanied by unknown modelling uncertainties, which have not been estimated by BMN. In Figure~\ref{fig:intensity_water}, we find agreement with the vertical intensities predicted over the entire range of depths by the {\sc mum} code, as well as the BMN and the \citet{Sinegovskaya:2000bv} calculations within our errors, although these do not estimate inclusive muon flux uncertainties. The parameterisation by \citet{Klimushin:2000cy} quotes a maximal 8\% error between the parameterisation and the numerical BMN result. This number applies to the deviation between a parameterisation used by the {\sc mum} code \citep{Sokalski:2000nb} and BMN, and is not a conservative error estimate of the underground intensity as recently quoted by the KM3NeT collaboration \citep{KM3NeT:2019jfa}.

The error bands of the \ddm{} and \sibyll{} calculations include only hadronic uncertainties in Figures~\ref{fig:intensity_true} and \ref{fig:comparison}. The dominant sources are pion and kaon production yields at small scattering angles. An additional, uncorrelated uncertainty of up to 20\% at large depths is expected from the limited knowledge of cosmic ray nucleon fluxes \citep{Evans:2016obt, Fedynitch:2018vfe,Dembinski:2017zsh}. Compared to the modelling uncertainty $\gtrsim12\%$ (for \ddm{}), the errors on the data appear much smaller. There is, therefore, a significant potential to obtain tighter constraints for muons up to tens of TeV. This result is of particular importance for the calibration of atmospheric neutrino flux calculations \citep{Sanuki:2006yd,Yanez:2021qd}. If the data can be cross-calibrated within the quoted systematic uncertainties, the high-energy neutrino flux may be known with a precision of $\sim10\%$, significantly below the $30-50\%$ of the present calculations of \citet{Barr:2006it} and \citet{Fedynitch:2012fs}. Consequently, a combination of surface muon and cosmic ray data can potentially add new constraints on proton and helium components in the energy ``gap'' between space-borne direct and indirect air-shower observations.

A secondary source of uncertainty is seasonal variations, which can account for up to an additional 2\%. Typically deep underground measurements are performed over multiple years such that the seasonal variations average out. However, seasonal variations can be used to study hadronic interactions and can serve as an additional source of constraints, as shown by, \textit{e.g.}~\citet{Grashorn:2009ey}. In addition, variations due to rock density or chemical composition can result in differences around 2\%, estimated by varying the density between $2.65$ and $2.84$ g cm$^{-3}$ in simulations with \mute{}.

For applications that depend on the theoretical knowledge of the underground muon flux, such as direct DM detectors, the result of our top-down calculations implies that the underground background rates are known with high certainty. The absence of empirical ingredients allows for accurate characterisation of the angular dependence, very rare events from near-horizontal zenith angles, and seasonal variations on short time scales.

\section{Conclusions}
\label{sec:conclusion}

We have developed a modern and lightweight open-source code, \mute{}, for precise calculations of underground muon fluxes, intensities, and rates. The scheme is a convolution of tabulated output from two state-of-the-art open-source codes, \mceq{} for surface flux calculations, and \proposal{} for underground transport. The properties or physical models can be varied independently for atmospheric cascades and underground propagation, allowing for a variety of possible applications.

We have assessed the energy ranges and parent meson fractions, confirming the previous BMN result on a subdominant prompt muon contribution at large depths. Prompt muons from decays of charm mesons and rare decays of unflavoured mesons contribute $\sim10\%$ of the vertical-equivalent intensity at depths below 12 km.w.e.\ according to the \sibyll{-2.3d} model. Given the large uncertainties of the cosmic ray flux and the overlap with contributions from neutrino-induced muons, discerning the prompt component through measurements of the muon rate is unlikely to succeed without much stronger constraints on cosmic ray fluxes and hadronic yields in the relevant phase space.

The calculated vertical-equivalent underground intensities have been compared to data. The overall description is very good within the hadronic uncertainties of the models. When not considering systematic uncertainties, the \ddm{} is better at shallow and medium depths, whereas the \sibyll{} interaction model seems better at very large depths. The data from different experiments are, in general, mutually compatible within the errors. Some observable shifts are likely related to systematic uncertainties in, \textit{e.g.}~the exact determination of the overburden.

In comparison with MH, our calculation has a different depth dependence but compatible errors. Within the individual depth ranges, data are more in line with our calculation. The MH fit is determined by the empirical function proposed by \citet{crouch1987} to represent a compilation of data from various sites and overburden types. Naturally, controlling the systematic uncertainties of a data compilation is very challenging and has likely lead to a bias that was not resolved by MH.

The calculations by BMN, which is the baseline for the {\sc mum} code used for underwater neutrino experiments, are fully contained by the error band of our \ddm{} calculation. This is remarkable since the hadronic and cosmic ray flux models have been parameterised using data from the 1980s or before. However, the error of 8\% quoted in an unrelated follow-up work is underestimated compared to our results, which have hadronic errors in the 15\% range, or more, if cosmic ray flux uncertainties are included.

The fact that the modelling uncertainties are considerably larger than the statistical uncertainties of the data is a significant result. If for some of the measurements, the systematic shifts can be cross-calibrated, the free parameters of the physics models can be constrained. Since \ddm{} and GSF are both data-driven models, adding high-precision and high-energy data will likely reduce the total error below 10\%. While the degeneracy between the hadronic model and the cosmic ray flux can not easily be resolved with just the underground intensity, adding more measurements on, \textit{e.g.}~seasonal variations or the underground charge ratio \citep{OPERA:2010cos}, can help break this degeneracy.

A cross-calibration of experimental systematics is beyond the scope of this work. It is, however, evident that even a single well-characterised underground intensity measurement has the potential to significantly constrain forward meson yields at projectile energies above TeV and the primary cosmic ray fluxes at the ``gap'' between direct and indirect measurements. We, therefore, encourage underground laboratories that have good control of systematics and sufficient statistics, such as Super-Kamiokande, to contribute a reference data set. As we extensively discuss, special attention should be given to zenith angle dependence. The added confidence would allow us to scrutinise the precision of the older data, opening the potential to push atmospheric muon and neutrino flux uncertainties in the multi-TeV range below 10\%. Smaller atmospheric flux uncertainties benefit the neutrino astronomy community, in particular in the determination of the astrophysical flux at lower energies, and for new physics searches.

\begin{acknowledgments}
We acknowledge the help of Michel Zampaolo and Luigi Mosca for the data of the Fréjus detector, as well as the \proposal{} developers, Jan Soedingrekso, Alexander Sandrock, Jean-Marco Alameddine, and Maximilian Sackel for their invaluable help. We gratefully acknowledge the support of Compute Canada (\url{www.computecanada.ca}) for providing the computing resources required to undertake this work. W.W. and M.-C.P. acknowledge the support from the McDonald Institute. A.F.\ acknowledges the hospitality within the group of Hiroyuki Sagawa at the ICRR, where he completed this work as JSPS International Research Fellow (JSPS KAKENHI Grant Number 19F19750).
\end{acknowledgments}

\bibliography{bibliography}

\begin{thebibliography}{}
\expandafter\ifx\csname natexlab\endcsname\relax\def\natexlab#1{#1}\fi
\providecommand{\url}[1]{\href{#1}{#1}}
\providecommand{\dodoi}[1]{doi:~\href{http://doi.org/#1}{\nolinkurl{#1}}}
\providecommand{\doeprint}[1]{\href{http://ascl.net/#1}{\nolinkurl{http://ascl.net/#1}}}
\providecommand{\doarXiv}[1]{\href{https://arxiv.org/abs/#1}{\nolinkurl{https://arxiv.org/abs/#1}}}

\bibitem[{Aartsen {et~al.}(2016{\natexlab{a}})}]{IceCube:2015wro}
Aartsen, M.~G., {et~al.} 2016{\natexlab{a}}, Astropart. Phys., 78, 1,
  \dodoi{10.1016/j.astropartphys.2016.01.006}

\bibitem[{Aartsen {et~al.}(2016{\natexlab{b}})}]{IceCube:2016umi}
---. 2016{\natexlab{b}}, Astrophys. J., 833, 3,
  \dodoi{10.3847/0004-637X/833/1/3}

\bibitem[{Aartsen {et~al.}(2017)}]{IceCube:2017roe}
---. 2017, Nature, 551, 596, \dodoi{10.1038/nature24459}

\bibitem[{Aartsen {et~al.}(2019)}]{IceCube:2019hmk}
---. 2019, Phys. Rev. D, 100, 082002, \dodoi{10.1103/PhysRevD.100.082002}

\bibitem[{Aartsen {et~al.}(2020)}]{IceCube:2020phf}
---. 2020, Phys. Rev. Lett., 125, 141801,
  \dodoi{10.1103/PhysRevLett.125.141801}

\bibitem[{Abbasi {et~al.}(2021)}]{IceCube:2020wum}
Abbasi, R., {et~al.} 2021, Phys. Rev. D, 104, 022002,
  \dodoi{10.1103/PhysRevD.104.022002}

\bibitem[{Abgrall {et~al.}(2016)}]{Abgrall:2015hmv}
Abgrall, N., {et~al.} 2016, Eur. Phys. J. C, 76, 84,
  \dodoi{10.1140/epjc/s10052-016-3898-y}

\bibitem[{Abramowicz {et~al.}(1991)Abramowicz, Levin, Levy, \&
  Maor}]{ABRAMOWICZ1991465}
Abramowicz, H., Levin, E., Levy, A., \& Maor, U. 1991, Physics Letters B, 269,
  465, \dodoi{https://doi.org/10.1016/0370-2693(91)90202-2}

\bibitem[{Abramowicz \& Levy(1997)}]{Abramowicz:1997ms}
Abramowicz, H., \& Levy, A. 1997.
\newblock \doarXiv{hep-ph/9712415}

\bibitem[{Adriani {et~al.}(2011)}]{PAMELA:2011mvy}
Adriani, O., {et~al.} 2011, Science, 332, 69, \dodoi{10.1126/science.1199172}

\bibitem[{Adriani {et~al.}(2019)}]{CALET:2019bmh}
---. 2019, Phys. Rev. Lett., 122, 181102,
  \dodoi{10.1103/PhysRevLett.122.181102}

\bibitem[{Agafonova {et~al.}(2010)}]{OPERA:2010cos}
Agafonova, N., {et~al.} 2010, Eur. Phys. J. C, 67, 25,
  \dodoi{10.1140/epjc/s10052-010-1284-8}

\bibitem[{Ageron {et~al.}(2020)}]{KM3NeT:2019jfa}
Ageron, M., {et~al.} 2020, Eur. Phys. J. C, 80, 99,
  \dodoi{10.1140/epjc/s10052-020-7629-z}

\bibitem[{Aglietta {et~al.}(1998)}]{LVD:1998lir}
Aglietta, M., {et~al.} 1998, Phys. Rev. D, 58, 092005,
  \dodoi{10.1103/PhysRevD.58.092005}

\bibitem[{Aglietta {et~al.}(1999)}]{LVD:1999khf}
---. 1999, Phys. Rev. D, 60, 112001, \dodoi{10.1103/PhysRevD.60.112001}

\bibitem[{Agostinelli {et~al.}(2003)}]{GEANT4:2002zbu}
Agostinelli, S., {et~al.} 2003, Nucl. Instrum. Meth. A, 506, 250,
  \dodoi{10.1016/S0168-9002(03)01368-8}

\bibitem[{Aguilar {et~al.}(2015)}]{AMS:2015tnn}
Aguilar, M., {et~al.} 2015, Phys. Rev. Lett., 114, 171103,
  \dodoi{10.1103/PhysRevLett.114.171103}

\bibitem[{Aharmim {et~al.}(2009)Aharmim, Ahmed, Andersen, Anthony, Barros,
  Beier, Bellerive, Beltran, Bergevin, Biller, Boudjemline, Boulay, Burritt,
  Cai, Chan, Chen, Chon, Cleveland, Cox-Mobrand, Currat, Dai, Dalnoki-Veress,
  Deng, Detwiler, Doe, Dosanjh, Doucas, Drouin, Duncan, Dunford, Elliott,
  Evans, Ewan, Farine, Fergani, Fleurot, Ford, Formaggio, Gagnon, Goon, Graham,
  Grant, Guillian, Habib, Hahn, Hallin, Hallman, Hargrove, Harvey, Hazama,
  Heeger, Heintzelman, Heise, Helmer, Hemingway, Henning, Hime, Howard, Howe,
  Huang, Jamieson, Jelley, Klein, Kos, Kr\"uger, Kraus, Krauss, Kutter, Kyba,
  Lange, Law, Lawson, Lesko, Leslie, Levine, Loach, Luoma, MacLellan, Majerus,
  Mak, Maneira, Marino, Martin, McCauley, McDonald, McGee, Mifflin, Miller,
  Monreal, Monroe, Noble, Oblath, Okada, O'Keeffe, Opachich, Gann, Oser, Ott,
  Peeters, Poon, Prior, Rielage, Robertson, Robertson, Rollin, Schwendener,
  Secrest, Seibert, Simard, Simpson, Sinclair, Skensved, Smith, Sonley,
  Steiger, Stonehill, Tagg, Te\ifmmode \check{s}\else
  \v{s}\fi{}i\ifmmode~\acute{c}\else \'{c}\fi{}, Tolich, Tsui, de~Water,
  VanDevender, Virtue, Waller, Waltham, Tseung, Wark, Watson, Wendland, West,
  Wilkerson, Wilson, Wouters, Wright, Yeh, Zhang, \& Zuber}]{sno2009}
Aharmim, B., Ahmed, S.~N., Andersen, T.~C., {et~al.} 2009, Phys. Rev. D, 80,
  012001, \dodoi{10.1103/PhysRevD.80.012001}

\bibitem[{Ahn {et~al.}(2009)Ahn, Engel, Gaisser, Lipari, \&
  Stanev}]{Ahn:2009wx}
Ahn, E.-J., Engel, R., Gaisser, T.~K., Lipari, P., \& Stanev, T. 2009, Phys.
  Rev. D, 80, 094003, \dodoi{10.1103/PhysRevD.80.094003}

\bibitem[{Aiello {et~al.}(2015)}]{NEMO:2014htj}
Aiello, S., {et~al.} 2015, Astropart. Phys., 66, 1,
  \dodoi{10.1016/j.astropartphys.2014.12.010}

\bibitem[{Allkofer {et~al.}(1985)Allkofer, Jokisch, Klemke, Oren, Uhr, Bella,
  \& Dau}]{Allkofer:1985ey}
Allkofer, O.~C., Jokisch, H., Klemke, G., {et~al.} 1985, Nucl. Phys. B, 259, 1,
  \dodoi{10.1016/0550-3213(86)90270-1}

\bibitem[{Alt {et~al.}(2007)}]{Alt:2006fr}
Alt, C., {et~al.} 2007, Eur. Phys. J., C49, 897,
  \dodoi{10.1140/epjc/s10052-006-0165-7}

\bibitem[{Alves~Jr. {et~al.}(2021)Alves~Jr., Reininghaus, Alameddine, Albrecht,
  Alvarez-Muniz, Arrabito, Baack, Bernlöhr, Bleicher, BREGEON, Carrere,
  Dembinski, Elfner, Elsässer, Engel, Hu, Fedynitch, Heck, Huege, Kampert,
  Karastathis, Nellen, Nöthe, Parello, Pierog, Pokrandt, Poctarev, Prechelt,
  Rhode, Riehn, Sackel, Sandrock, Sampathkumar, Schmelling, Schmidt, Sigl,
  Soedingrekso, Spaan, Xu, Ammerman-Yebra, Zas, \& Ulrich}]{CORSIKA8:20212c}
Alves~Jr., A.~A., Reininghaus, M., Alameddine, J.-M., {et~al.} 2021, PoS,
  ICRC2021, 474, \dodoi{10.22323/1.395.0474}

\bibitem[{Ambrosio {et~al.}(1995)Ambrosio, Antolini, Auriemma, Baker, Baldini,
  Barbarino, Barish, Battistoni, Bellotti, Bemporad, Bernardini, Bilokon, Bisi,
  Bloise, Bower, Bussino, Cafagna, Calicchio, Campana, Carboni, Castellano,
  Cecchini, Cei, Celio, Chiarella, Corona, Coutu, De~Cataldo, Dekhissi,
  De~Marzo, De~Mitri, De~Vincenzi, Di~Credico, Erriquez, Favuzzi, Forti, Fusco,
  Giacomelli, Giannini, Giglietto, Grassi, Grillo, Guarino, Guarnaccia,
  Gustavino, Habig, Hanson, Hawthorne, Heinz, Hong, Iarocci, Katsavounidis,
  Kearns, Kyriazopoulou, Lamanna, Lane, Levin, Lipari, Liu, Longley, Longo, Lu,
  Ludlam, Mancarella, Mandrioli, Margiotta-Neri, Marini, Martello,
  Marzari-Chiesa, Mazziotta, Michael, Mikheyev, Miller, Mittelbrunn, Monacelli,
  Montaruli, Monteno, Mufson, Musser, Nicol\'o, Nolty, Okada, Orth, Osteria,
  Palamara, Parlati, Patera, Patrizii, Pazzi, Peck, Petrera, Pignatano,
  Pistilli, Popa, Rain\'o, Reynoldson, Ronga, Sanzgiri, Sartogo, Satriano,
  Satta, Scapparone, Scholberg, Sciubba, Serra-Lugaresi, Severi, Sitta,
  Spinelli, Spinetti, Spurio, Steinberg, Stone, Sulak, Surdo, Tarl\'e, Tassoni,
  Togo, Valente, Walter, \& Webb}]{macro1995}
Ambrosio, M., Antolini, R., Auriemma, G., {et~al.} 1995, Phys. Rev. D, 52,
  3793, \dodoi{10.1103/PhysRevD.52.3793}

\bibitem[{An {et~al.}(2019)}]{DAMPE:2019gys}
An, Q., {et~al.} 2019, Sci. Adv., 5, eaax3793, \dodoi{10.1126/sciadv.aax3793}

\bibitem[{{Andreyev} {et~al.}(1987){Andreyev}, {Gurentsov}, \&
  {Kogai}}]{baksan1987}
{Andreyev}, Y.~M., {Gurentsov}, V.~I., \& {Kogai}, I.~M. 1987, in International
  Cosmic Ray Conference, Vol.~6, International Cosmic Ray Conference, 200

\bibitem[{Anticic {et~al.}(2010)}]{Anticic:2010yg}
Anticic, T., {et~al.} 2010, Eur. Phys. J. C, 68, 1,
  \dodoi{10.1140/epjc/s10052-010-1328-0}

\bibitem[{Apel {et~al.}(2012)}]{Apel:2012tda}
Apel, W.~D., {et~al.} 2012, Astropart. Phys., 36, 183,
  \dodoi{10.1016/j.astropartphys.2012.05.023}

\bibitem[{Bai {et~al.}(2003)}]{Bai:2003mn}
Bai, X., {et~al.} 2003, in {28th International Cosmic Ray Conference},
  1373--1376

\bibitem[{Barr {et~al.}(2006)Barr, Gaisser, Robbins, \& Stanev}]{Barr:2006it}
Barr, G.~D., Gaisser, T.~K., Robbins, S., \& Stanev, T. 2006, Phys. Rev. D, 74,
  094009, \dodoi{10.1103/PhysRevD.74.094009}

\bibitem[{Barrett {et~al.}(1952)Barrett, Bollinger, Cocconi, Eisenberg, \&
  Greisen}]{Barrett:1952woo}
Barrett, P.~H., Bollinger, L.~M., Cocconi, G., Eisenberg, Y., \& Greisen, K.
  1952, Rev. Mod. Phys., 24, 133, \dodoi{10.1103/RevModPhys.24.133}

\bibitem[{Battistoni {et~al.}(2015)}]{Battistoni:2015epi}
Battistoni, G., {et~al.} 2015, Annals Nucl. Energy, 82, 10,
  \dodoi{10.1016/j.anucene.2014.11.007}

\bibitem[{Bazzotti(2009)}]{Bazzotti:2009sy}
Bazzotti, M. 2009.
\newblock \doarXiv{0911.3055}

\bibitem[{Becker~Tjus \& Merten(2020)}]{BeckerTjus:2020xzg}
Becker~Tjus, J., \& Merten, L. 2020, Phys. Rept., 872, 1,
  \dodoi{10.1016/j.physrep.2020.05.002}

\bibitem[{Berger {et~al.}(1989)Berger, Fr\"ohlich, M\"onch, Nisius, Raupach,
  Schleper, Benadjal, Blum, Bourdarious, Dudelzak, Eschstruth, Jullian,
  Lalanne, Laplanche, Longuemare, Paulot, Perdereau, Roy, Szklarz, Behr,
  Degrange, Minet, Nguyen-Khac, Serri, Tisserant, Tripp, Arpesella, Bareyre,
  Barloutaud, Borg, Chardin, Ernwein, Glicenstein, Mosca, Moscoso, Becker,
  Becker, Daum, Demski, Jacobi, Kuznik, Mayer, Meyer, M\"oller, Schubnell,
  Seyffert, Wei, \& Wintgen}]{frejus1989}
Berger, C., Fr\"ohlich, M., M\"onch, H., {et~al.} 1989, Phys. Rev. D, 40, 2163,
  \dodoi{10.1103/PhysRevD.40.2163}

\bibitem[{Bergmann {et~al.}(2007)Bergmann, Engel, Heck, Kalmykov, Ostapchenko,
  Pierog, Thouw, \& Werner}]{Bergmann:2006yz}
Bergmann, T., Engel, R., Heck, D., {et~al.} 2007, Astropart. Phys., 26, 420,
  \dodoi{10.1016/j.astropartphys.2006.08.005}

\bibitem[{Brodsky {et~al.}(1980)Brodsky, Hoyer, Peterson, \&
  Sakai}]{Brodsky:1980pb}
Brodsky, S.~J., Hoyer, P., Peterson, C., \& Sakai, N. 1980, Phys. Lett. B, 93,
  451, \dodoi{10.1016/0370-2693(80)90364-0}

\bibitem[{Bugaev {et~al.}(1998)Bugaev, Misaki, Naumov, Sinegovskaya,
  Sinegovsky, \& Takahashi}]{Bugaev:1998bi}
Bugaev, E.~V., Misaki, A., Naumov, V.~A., {et~al.} 1998, Phys. Rev. D, 58,
  054001, \dodoi{10.1103/PhysRevD.58.054001}

\bibitem[{Bugaev {et~al.}(1989)Bugaev, Naumov, Sinegovsky, \&
  Zaslavskaya}]{Bugaev:1989we}
Bugaev, E.~V., Naumov, V.~A., Sinegovsky, S.~I., \& Zaslavskaya, E.~S. 1989,
  Nuovo Cim. C, 12, 41, \dodoi{10.1007/BF02509070}

\bibitem[{Chirkin \& Rhode(2004)}]{Chirkin:2004hz}
Chirkin, D., \& Rhode, W. 2004.
\newblock \doarXiv{hep-ph/0407075}

\bibitem[{{Crouch}(1987)}]{crouch1987}
{Crouch}, M. 1987, in International Cosmic Ray Conference, Vol.~6,
  International Cosmic Ray Conference, 165

\bibitem[{Dembinski {et~al.}(2018)Dembinski, Engel, Fedynitch, Gaisser, Riehn,
  \& Stanev}]{Dembinski:2017zsh}
Dembinski, H.~P., Engel, R., Fedynitch, A., {et~al.} 2018, PoS, ICRC2017, 533,
  \dodoi{10.22323/1.301.0533}

\bibitem[{Donini {et~al.}(2019)Donini, Palomares-Ruiz, \&
  Salvado}]{Donini:2018tsg}
Donini, A., Palomares-Ruiz, S., \& Salvado, J. 2019, Nature Phys., 15, 37,
  \dodoi{10.1038/s41567-018-0319-1}

\bibitem[{Dunsch {et~al.}(2019)Dunsch, Soedingrekso, Sandrock, Meier, Menne, \&
  Rhode}]{dunsch_2018_proposal_improvements}
Dunsch, M., Soedingrekso, J., Sandrock, A., {et~al.} 2019, Computer Physics
  Communications, 242, 132, \dodoi{10.1016/j.cpc.2019.03.021}

\bibitem[{Elbert {et~al.}(1983)Elbert, Gaisser, \& Stanev}]{Elbert:1982xj}
Elbert, J.~W., Gaisser, T.~K., \& Stanev, T. 1983, Phys. Rev. D, 27, 1448,
  \dodoi{10.1103/PhysRevD.27.1448}

\bibitem[{Engel {et~al.}(2019)Engel, Heck, Huege, Pierog, Reininghaus, Riehn,
  Ulrich, Unger, \& Veberi\v{c}}]{Engel:2018akg}
Engel, R., Heck, D., Huege, T., {et~al.} 2019, Comput. Softw. Big Sci., 3, 2,
  \dodoi{10.1007/s41781-018-0013-0}

\bibitem[{Evans {et~al.}(2017)Evans, Gamez, Porzio, S\"oldner-Rembold, \&
  Wren}]{Evans:2016obt}
Evans, J., Gamez, D.~G., Porzio, S.~D., S\"oldner-Rembold, S., \& Wren, S.
  2017, Phys. Rev. D, 95, 023012, \dodoi{10.1103/PhysRevD.95.023012}

\bibitem[{Fedynitch(2015)}]{Fedynitch:2015kcn}
Fedynitch, A. 2015, PhD thesis, KIT, Karlsruhe, Dept. Phys.,
  \dodoi{10.5445/IR/1000055433}

\bibitem[{Fedynitch {et~al.}(2012)Fedynitch, Becker~Tjus, \&
  Desiati}]{Fedynitch:2012fs}
Fedynitch, A., Becker~Tjus, J., \& Desiati, P. 2012, Phys. Rev. D, 86, 114024,
  \dodoi{10.1103/PhysRevD.86.114024}

\bibitem[{Fedynitch {et~al.}(2018)Fedynitch, Dembinski, Engel, Gaisser, Riehn,
  \& Stanev}]{Fedynitch:2018vfe}
Fedynitch, A., Dembinski, H., Engel, R., {et~al.} 2018, PoS, ICRC2017, 1019,
  \dodoi{10.22323/1.301.1019}

\bibitem[{Fedynitch \& Huber(2021)}]{Fedynitch:2021ks}
Fedynitch, A., \& Huber, M. 2021, PoS, ICRC2021, 1227,
  \dodoi{10.22323/1.395.1227}

\bibitem[{Fedynitch {et~al.}(2019)Fedynitch, Riehn, Engel, Gaisser, \&
  Stanev}]{Fedynitch:2018cbl}
Fedynitch, A., Riehn, F., Engel, R., Gaisser, T.~K., \& Stanev, T. 2019, Phys.
  Rev. D, 100, 103018, \dodoi{10.1103/PhysRevD.100.103018}

\bibitem[{Formaggio \& Martoff(2004)}]{Formaggio:2004mm}
Formaggio, J.~A., \& Martoff, C. 2004, Annual Review of Nuclear and Particle
  Science, 54, 361, \dodoi{10.1146/annurev.nucl.54.070103.181248}

\bibitem[{Gaisser(2012)}]{Gaisser:2012zz}
Gaisser, T.~K. 2012, Astropart. Phys., 35, 801,
  \dodoi{10.1016/j.astropartphys.2012.02.010}

\bibitem[{Gaisser {et~al.}(2016)Gaisser, Engel, \& Resconi}]{Gaisser:2016uoy}
Gaisser, T.~K., Engel, R., \& Resconi, E. 2016, {Cosmic Rays and Particle
  Physics}: {2nd Edition} (Cambridge University Press)

\bibitem[{Gaisser {et~al.}(1995)Gaisser, Halzen, \& Stanev}]{Gaisser:1994yf}
Gaisser, T.~K., Halzen, F., \& Stanev, T. 1995, Phys. Rept., 258, 173,
  \dodoi{10.1016/0370-1573(95)00003-Y}

\bibitem[{Gaisser \& Honda(2002)}]{Gaisser:2002jj}
Gaisser, T.~K., \& Honda, M. 2002, Ann. Rev. Nucl. Part. Sci., 52, 153,
  \dodoi{10.1146/annurev.nucl.52.050102.090645}

\bibitem[{Gaisser {et~al.}(2020)Gaisser, Soldin, Crossman, \&
  Fedynitch}]{Gaisser:2019xlw}
Gaisser, T.~K., Soldin, D., Crossman, A., \& Fedynitch, A. 2020, PoS, ICRC2019,
  893, \dodoi{10.22323/1.358.0893}

\bibitem[{Gaisser {et~al.}(2013)Gaisser, Stanev, \& Tilav}]{Gaisser:2013bla}
Gaisser, T.~K., Stanev, T., \& Tilav, S. 2013, Front. Phys. (Beijing), 8, 748,
  \dodoi{10.1007/s11467-013-0319-7}

\bibitem[{Garzelli {et~al.}(2015)Garzelli, Moch, \& Sigl}]{Garzelli:2015psa}
Garzelli, M.~V., Moch, S., \& Sigl, G. 2015, JHEP, 10, 115,
  \dodoi{10.1007/JHEP10(2015)115}

\bibitem[{Gondolo {et~al.}(1996)Gondolo, Ingelman, \& Thunman}]{Gondolo:1995fq}
Gondolo, P., Ingelman, G., \& Thunman, M. 1996, Astropart. Phys., 5, 309,
  \dodoi{10.1016/0927-6505(96)00033-3}

\bibitem[{Grashorn {et~al.}(2010)Grashorn, de~Jong, Goodman, Habig, Marshak,
  Mufson, Osprey, \& Schreiner}]{Grashorn:2009ey}
Grashorn, E.~W., de~Jong, J.~K., Goodman, M.~C., {et~al.} 2010, Astropart.
  Phys., 33, 140, \dodoi{10.1016/j.astropartphys.2009.12.006}

\bibitem[{Heck {et~al.}(1998)Heck, Knapp, Capdevielle, Schatz, \&
  Thouw}]{Heck:1998vt}
Heck, D., Knapp, J., Capdevielle, J.~N., Schatz, G., \& Thouw, T. 1998

\bibitem[{Hoerandel(2003)}]{Hoerandel:2002yg}
Hoerandel, J.~R. 2003, Astropart. Phys., 19, 193,
  \dodoi{10.1016/S0927-6505(02)00198-6}

\bibitem[{Honda {et~al.}(2007)Honda, Kajita, Kasahara, Midorikawa, \&
  Sanuki}]{Honda:2006qj}
Honda, M., Kajita, T., Kasahara, K., Midorikawa, S., \& Sanuki, T. 2007, Phys.
  Rev. D, 75, 043006, \dodoi{10.1103/PhysRevD.75.043006}

\bibitem[{Honda {et~al.}(2019)Honda, Sajjad~Athar, Kajita, Kasahara, \&
  Midorikawa}]{Honda:2019ymh}
Honda, M., Sajjad~Athar, M., Kajita, T., Kasahara, K., \& Midorikawa, S. 2019,
  Phys. Rev. D, 100, 123022, \dodoi{10.1103/PhysRevD.100.123022}

\bibitem[{{Hoshina} \& {Tanaka}(2012)}]{2012AGUFM.P23D..03H}
{Hoshina}, K., \& {Tanaka}, H. 2012, in AGU Fall Meeting Abstracts, Vol. 2012,
  P23D--03

\bibitem[{Illana {et~al.}(2011)Illana, Lipari, Masip, \&
  Meloni}]{Illana:2010gh}
Illana, J.~I., Lipari, P., Masip, M., \& Meloni, D. 2011, Astropart. Phys., 34,
  663, \dodoi{10.1016/j.astropartphys.2011.01.001}

\bibitem[{Kelner(1998)}]{Kelner:1998mh}
Kelner, S.~R. 1998, Phys. Atom. Nucl., 61, 448

\bibitem[{Kelner {et~al.}(1995)Kelner, Kokoulin, \& Petrukhin}]{Kelner:1995hu}
Kelner, S.~R., Kokoulin, R.~P., \& Petrukhin, A.~A. 1995, {About cross-section
  for high-energy muon bremsstrahlung}, Tech. Rep. FPRINT-95-36

\bibitem[{Klimushin {et~al.}(2001)Klimushin, Bugaev, \&
  Sokalski}]{Klimushin:2000cy}
Klimushin, S.~I., Bugaev, E.~V., \& Sokalski, I.~A. 2001, Phys. Rev. D, 64,
  014016, \dodoi{10.1103/PhysRevD.64.014016}

\bibitem[{Kochanov {et~al.}(2019)Kochanov, Morozova, Sinegovskaya, \&
  Sinegovsky}]{Kochanov:2019yvx}
Kochanov, A.~A., Morozova, A.~D., Sinegovskaya, T.~S., \& Sinegovsky, S.~I.
  2019, J. Phys. Conf. Ser., 1181, 012054,
  \dodoi{10.1088/1742-6596/1181/1/012054}

\bibitem[{Kochanov {et~al.}(2008)Kochanov, Sinegovskaya, \&
  Sinegovsky}]{Kochanov:2008pt}
Kochanov, A.~A., Sinegovskaya, T.~S., \& Sinegovsky, S.~I. 2008, Astropart.
  Phys., 30, 219, \dodoi{10.1016/j.astropartphys.2008.09.008}

\bibitem[{Koehne {et~al.}(2013)Koehne, Frantzen, Schmitz, Fuchs, Rhode,
  Chirkin, \& Tjus}]{koehne2013proposal}
Koehne, J.-H., Frantzen, K., Schmitz, M., {et~al.} 2013, Computer Physics
  Communications, 184, 2070, \dodoi{10.1016/j.cpc.2013.04.001}

\bibitem[{Kokoulin(1999)}]{Kokoulin:1999bn}
Kokoulin, R.~P. 1999, Nucl. Phys. B Proc. Suppl., 70, 475,
  \dodoi{10.1016/S0920-5632(98)00475-7}

\bibitem[{Kozynets {et~al.}(2021)Kozynets, Fedynitch, \&
  Koskinen}]{Kozynets:2021LK}
Kozynets, T., Fedynitch, A., \& Koskinen, D.~J. 2021, PoS, ICRC2021, 1209,
  \dodoi{10.22323/1.395.1209}

\bibitem[{Kudryavtsev(2009)}]{Kudryavtsev:2008qh}
Kudryavtsev, V.~A. 2009, Comput. Phys. Commun., 180, 339,
  \dodoi{10.1016/j.cpc.2008.10.013}

\bibitem[{Lagutin {et~al.}(2004)Lagutin, Tyumentsev, \&
  Yushkov}]{Lagutin:2004ka}
Lagutin, A.~A., Tyumentsev, A.~G., \& Yushkov, A.~V. 2004, J. Phys. G, 30, 573,
  \dodoi{10.1088/0954-3899/30/5/003}

\bibitem[{Lagutin \& Yushkov(2006)}]{Lagutin:2006kv}
Lagutin, A.~A., \& Yushkov, A.~V. 2006, Phys. Atom. Nucl., 69, 460,
  \dodoi{10.1134/S1063778806030094}

\bibitem[{Lipari \& Stanev(1991)}]{Lipari:1991ut}
Lipari, P., \& Stanev, T. 1991, Phys. Rev. D, 44, 3543,
  \dodoi{10.1103/PhysRevD.44.3543}

\bibitem[{Manukovsky {et~al.}(2016)Manukovsky, Ryazhskaya, Sobolevsky, \&
  Yudin}]{Manukovsky:2016fcn}
Manukovsky, K.~V., Ryazhskaya, O.~G., Sobolevsky, N.~M., \& Yudin, A.~V. 2016,
  Phys. Atom. Nucl., 79, 631, \dodoi{10.1134/S106377881603011X}

\bibitem[{Mascaretti {et~al.}(2020)Mascaretti, Blasi, \&
  Evoli}]{Mascaretti:2019mnk}
Mascaretti, C., Blasi, P., \& Evoli, C. 2020, Astropart. Phys., 114, 22,
  \dodoi{10.1016/j.astropartphys.2019.06.002}

\bibitem[{Matsuno {et~al.}(1984)}]{Matsuno:1984kq}
Matsuno, S., {et~al.} 1984, Phys. Rev. D, 29, 1, \dodoi{10.1103/PhysRevD.29.1}

\bibitem[{Mei \& Hime(2006)}]{Mei:2005gm}
Mei, D., \& Hime, A. 2006, Phys. Rev. D, 73, 053004,
  \dodoi{10.1103/PhysRevD.73.053004}

\bibitem[{Nikolsky {et~al.}(1984)Nikolsky, Stamenov, \&
  Ushev}]{Nikolsky:1984emk}
Nikolsky, S.~I., Stamenov, I.~N., \& Ushev, S.~Z. 1984, Sov. Phys. JETP, 60, 10

\bibitem[{Ostapchenko(2011)}]{Ostapchenko:2010vb}
Ostapchenko, S. 2011, Phys. Rev., D83, 014018,
  \dodoi{10.1103/PhysRevD.83.014018}

\bibitem[{{Peters}(1961)}]{1961NCim...22..800P}
{Peters}, B. 1961, Il Nuovo Cimento, 22, 800, \dodoi{10.1007/BF02783106}

\bibitem[{Pierog {et~al.}(2015)Pierog, Karpenko, Katzy, Yatsenko, \&
  Werner}]{Pierog:2013ria}
Pierog, T., Karpenko, I., Katzy, J.~M., Yatsenko, E., \& Werner, K. 2015, Phys.
  Rev., C92, 034906, \dodoi{10.1103/PhysRevC.92.034906}

\bibitem[{Riehn {et~al.}(2020)Riehn, Engel, Fedynitch, Gaisser, \&
  Stanev}]{Engel:2019dsg}
Riehn, F., Engel, R., Fedynitch, A., Gaisser, T.~K., \& Stanev, T. 2020, Phys.
  Rev. D, 102, 063002, \dodoi{10.1103/PhysRevD.102.063002}

\bibitem[{Roesler {et~al.}(2000)Roesler, Engel, \& Ranft}]{dpmjetIII}
Roesler, S., Engel, R., \& Ranft, J. 2000, in {Advanced Monte Carlo for
  radiation physics, particle transport simulation and applications.
  Proceedings, Conference, MC2000, Lisbon, Portugal, October 23-26, 2000},
  1033--1038, \dodoi{10.1007/978-3-642-18211-2_166}

\bibitem[{Rossi(1952)}]{Rossi:99081}
Rossi, B.~B. 1952, {High-energy particles}, Prentice-Hall physics series (New
  York, NY: Prentice-Hall).
\newblock \url{https://cds.cern.ch/record/99081}

\bibitem[{Sandrock {et~al.}(2020)Sandrock, Kokoulin, \&
  Petrukhin}]{Sandrock:2020zaf}
Sandrock, A., Kokoulin, R.~P., \& Petrukhin, A.~A. 2020, J. Phys. Conf. Ser.,
  1690, 012005, \dodoi{10.1088/1742-6596/1690/1/012005}

\bibitem[{Sanuki {et~al.}(2007)Sanuki, Honda, Kajita, Kasahara, \&
  Midorikawa}]{Sanuki:2006yd}
Sanuki, T., Honda, M., Kajita, T., Kasahara, K., \& Midorikawa, S. 2007, Phys.
  Rev. D, 75, 043005, \dodoi{10.1103/PhysRevD.75.043005}

\bibitem[{Sciutto(1999)}]{Sciutto:1999jh}
Sciutto, S.~J. 1999, \dodoi{10.13140/RG.2.2.12566.40002}

\bibitem[{Sinegovskaya {et~al.}(2015)Sinegovskaya, Morozova, \&
  Sinegovsky}]{Sinegovskaya:2014pia}
Sinegovskaya, T.~S., Morozova, A.~D., \& Sinegovsky, S.~I. 2015, Phys. Rev. D,
  91, 063011, \dodoi{10.1103/PhysRevD.91.063011}

\bibitem[{Sinegovskaya \& Sinegovsky(2001)}]{Sinegovskaya:2000bv}
Sinegovskaya, T.~S., \& Sinegovsky, S.~I. 2001, Phys. Rev. D, 63, 096004,
  \dodoi{10.1103/PhysRevD.63.096004}

\bibitem[{Sokalski {et~al.}(2001)Sokalski, Bugaev, \&
  Klimushin}]{Sokalski:2000nb}
Sokalski, I.~A., Bugaev, E.~V., \& Klimushin, S.~I. 2001, Phys. Rev. D, 64,
  074015, \dodoi{10.1103/PhysRevD.64.074015}

\bibitem[{Sokalski {et~al.}(2002)Sokalski, Bugaev, \&
  Klimushin}]{Sokalski:2002dk}
Sokalski, I.~A., Bugaev, E.~V., \& Klimushin, S.~I. 2002, in {2nd Workshop on
  Methodical Aspects of Underwater/Ice Neutrino Telescopes}, 7--14.
\newblock \doarXiv{hep-ph/0201122}

\bibitem[{Standard~Atmosphere(1976)}]{us_std_atmosphere}
Standard~Atmosphere, U.~S. 1976, US Gov. Print. Off., Washington, DC

\bibitem[{Stettner(2020)}]{Stettner:2019tok}
Stettner, J. 2020, PoS, ICRC2019, 1017, \dodoi{10.22323/1.358.1017}

\bibitem[{Volkova \& Zatsepin(1983)}]{Volkova:1983yf}
Volkova, L.~v., \& Zatsepin, G.~t. 1983, Sov. J. Nucl. Phys., 37, 212

\bibitem[{Woodley(2021)}]{william_woodley_2021_5803936}
Woodley, W. 2021, wjwoodley/mute: MUTE 1.0.1, 1.0.1,  Zenodo,
  \dodoi{10.5281/zenodo.5803936}

\bibitem[{Y\'a\~nez {et~al.}(2020)Y\'a\~nez, Fedynitch, \&
  Montgomery}]{Yanez:2019bnw}
Y\'a\~nez, J.-P., Fedynitch, A., \& Montgomery, T. 2020, PoS, ICRC2019, 881,
  \dodoi{10.22323/1.358.0881}

\bibitem[{Yanez \& Fedynitch(2021)}]{Yanez:2021qd}
Yanez, J.-P., \& Fedynitch, A. 2021, PoS, ICRC2021, 1149,
  \dodoi{10.22323/1.395.1149}

\bibitem[{Zatsepin \& Sokolskaya(2006)}]{Zatsepin:2006ci}
Zatsepin, V.~I., \& Sokolskaya, N.~V. 2006, Astron. Astrophys., 458, 1,
  \dodoi{10.1051/0004-6361:20065108}

\bibitem[{Zyla {et~al.}(2020)}]{pdg2020}
Zyla, P., {et~al.} 2020, PTEP, 2020, 083C01, \dodoi{10.1093/ptep/ptaa104}

\end{thebibliography}
\bibliographystyle{aasjournal}

\begin{appendix}

\section{Dependence on the cosmic ray and hadronic model}
\label{app:primary_flux_and_hardonic_models}
\begin{figure}
\centering
\includegraphics[width=.49\columnwidth]{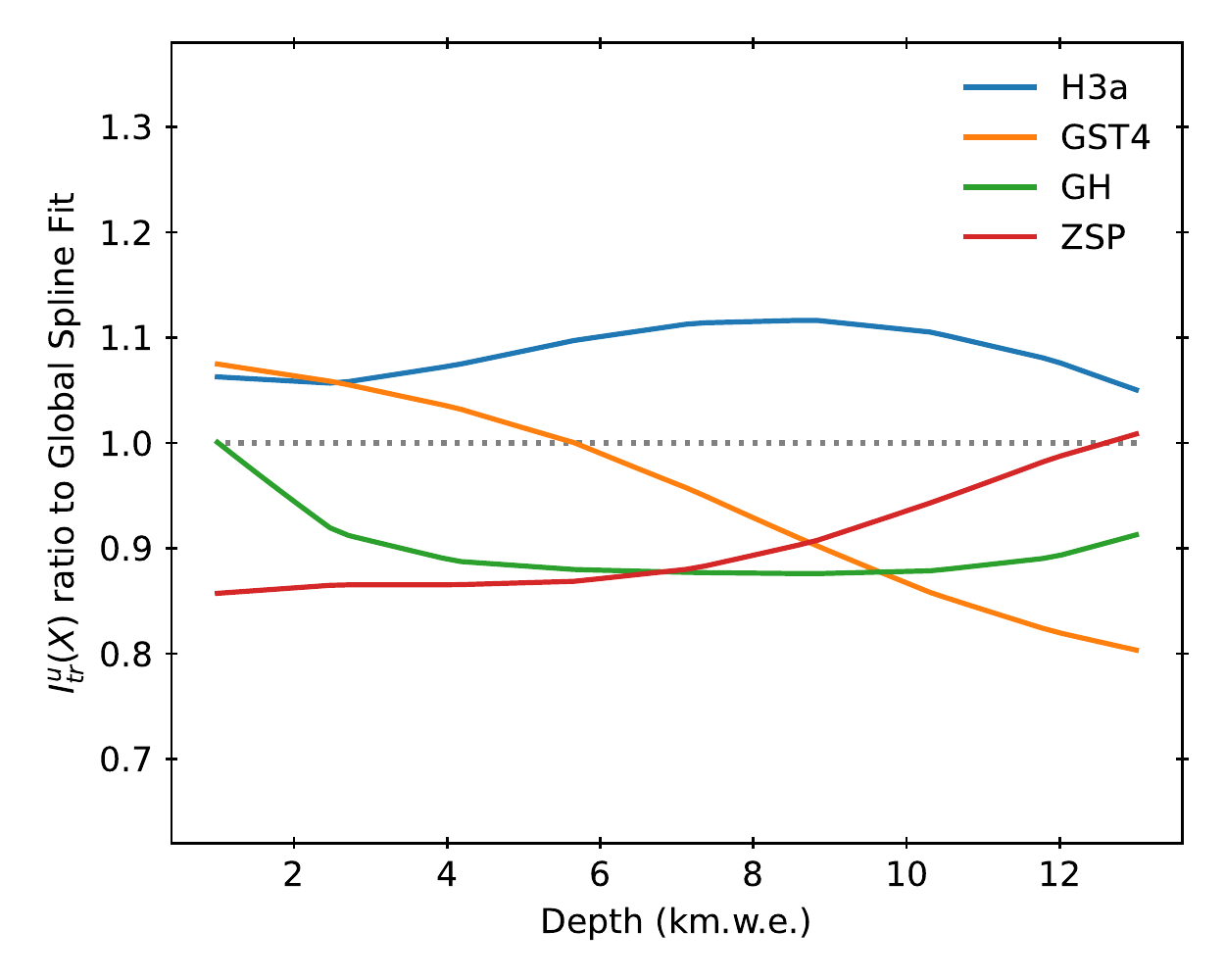}
\includegraphics[width=.49\columnwidth]{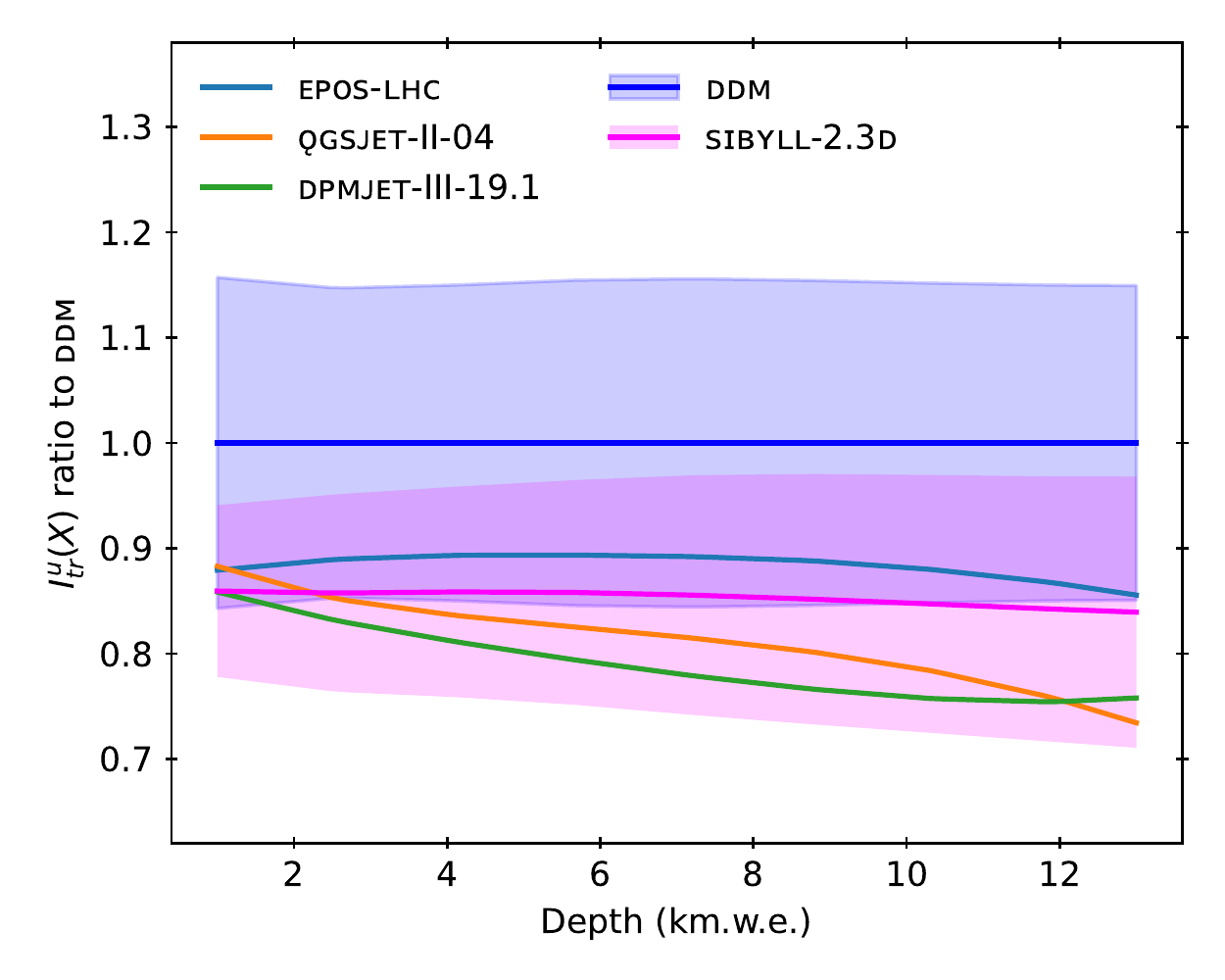}
\caption{Ratio of the true underground intensity for different choices of cosmic ray flux models (left) and hadronic interaction models (right) with cosmic ray flux model fixed to GSF. The interaction models are referenced in Section~\ref{sec:surface_models}. The primary models are parameterisations of data: H3a \citep{Gaisser:2012zz}, GST 4-generation \citep{Gaisser:2013bla}, GH \citep{Gaisser:2002jj}, and ZSP \citep{Zatsepin:2006ci,PAMELA:2011mvy}.}
\label{fig:ratio_pflux_models}
\end{figure}

We assess the impact of the cosmic ray flux model on the true vertical intensity in the left panel of Figure~\ref{fig:ratio_pflux_models}, and compare it to the impact from varying the hadronic interaction model in the right panel of the same figure. The nucleon fluxes, including the uncertainties of GSF, are shown in Figure 1 of \cite{Fedynitch:2018vfe}. Most of the differences at large depths are related to the composition modelling of the knee. The light-vs-heavy knee discussion has been brought forward by KASCADE and IceTop measurements that indicate a light, proton-rich composition at around PeV energies \citep{Apel:2012tda,IceCube:2019hmk}. This scenario is na\"{i}vely unexpected if a single dominant population of galactic cosmic ray accelerators cuts off at maximal rigidity, leading to Peters cycles \citep{1961NCim...22..800P}. The GST model \citep{Gaisser:2013bla} has an early cut-off and a softer spectrum in the proton component, which is reflected in a drop of intensities over 20\% compared to GSF. In principle, the underground intensities can help to fit the correct composition of cosmic rays at the knee if hadronic uncertainties and experimental systematics are carefully taken into account. However, at the moment, the hadronic uncertainties are sufficiently large to also cover the cosmic ray uncertainty.

The right panel of Figure~\ref{fig:ratio_pflux_models} demonstrates that our results are not significantly impacted by the choice of the interaction model. The \bartol{} errors applied to \sibyll{} entirely cover the predictions of the other models, whereas \dpmjet{} and \qgsjet{} drop out of the \ddm{} band. Both figures clearly demonstrate that there is an ambiguity between the cosmic ray and the hadronic model. The vertical intensity is too featureless for any reasonable separation between those. Hence, only the combination can be potentially constrained, {\it i.e.} the combination of \qgsjet{} with GST-4 would be in tension with Baksan and LVD at large depths ({\it cf.}~Figure~\ref{fig:comparison}).

\section{Impact of the vertical-equivalent approximation on the interpretation of data}
\label{app:vertical_equivalent_approximation}

\begin{figure}
\centering
\noindent\includegraphics[width=\columnwidth]{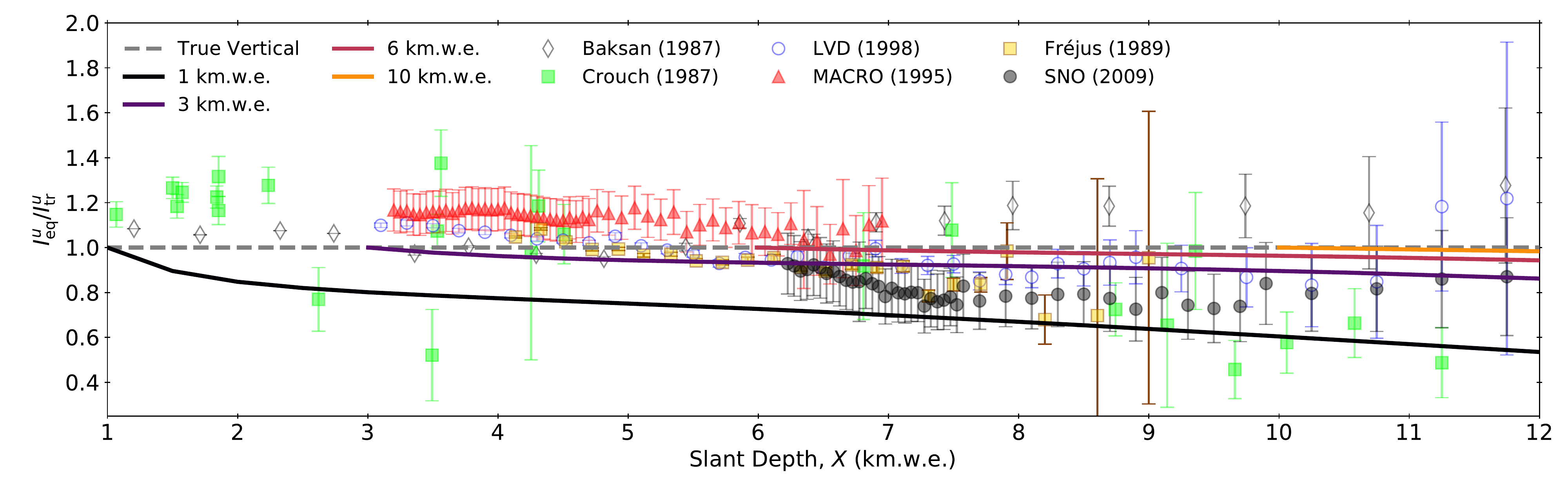}
\caption{The ratio of the vertical-equivalent intensity to the true vertical intensity against slant depth for various initial vertical depths as calculated by \mute{} for standard rock using \ddm{} and GSF. Each curve starts at the vertical depth indicated in the legend for $\theta=0$. All experimental data are referenced in Figure \ref{fig:intensity_true}. \label{fig:vertical_comparison}}
\end{figure}

Figure \ref{fig:vertical_comparison} shows that the deviation of vertical-equivalent intensity from true vertical intensity is worse for shallower vertical depths and deeper slant depths, as already shown in Figure \ref{fig:deviation}. The underground muon intensity data are seen to follow the trend of this deviation, as they are presented by the experiments in terms of vertical-equivalent intensities. The importance of the accurate treatment of the zenith angle dependence can be seen in comparison with Figure~\ref{fig:ratio_pflux_models}, where similar subtle tilts can be induced by the physical models. For this reason, we stress the importance of providing double-differential data in addition to using the vertical-equivalent approximation.

\end{appendix}

\end{document}